\documentclass[longauth,useAMS,usenatbib]{mn2e} 
\usepackage{times,amsmath,subfigure}
\usepackage{graphicx,natbib,float}
\usepackage{txfonts}
\def\aap{A\&A}                
\def\apjl{ApJ}                
\def\apjs{ApJS}               

\def\amm{$\mathrm{NH_3}$}
\def\hcop{$\mathrm{HCO^+}$}
\def\hh13cop{$\mathrm{H^{13}CO^+}$}
\def\n2hp{$\mathrm{N_2H^+}$}
\def\c18o{$\mathrm{C^{18}O}$}
\def\mecn{$\mathrm{CH_3CN}$}
\def\h2co{$\mathrm{H_2CO}$}
\def\arcsec{$''$}
\def\arcmin{$'$}
\def\um{$\mu$m}
\def\kms{$\mathrm{km~s^{-1}}$}
\def\hiir{UC H{\sc ii}~region}
\def\degr{$^\circ$}
\def\msun{$M_\odot$}

\def\thco{$\mathrm{^{13}CO}$}
\def\meth{$\mathrm{CH_3OH}$}

   \title[Sequential star formation near W\,48A]{A {\em Herschel} and BIMA study of the sequential star formation near the W\,48A H{\sc ii} region\thanks{{\em Herschel} is an ESA space observatory with science instruments provided by European-led Principal Investigator consortia and with important participation from NASA. This work is partially based on observations carried out with the IRAM 30m Telescope. IRAM is supported by INSU/CNRS (France), MPG (Germany) and IGN (Spain).} }

   \author[K.~L.~J. Rygl et al.]
{\large K.~L.~J. Rygl$^{1,2}$\thanks{E-mail:krygl@rssd.esa.int}, S. Goedhart$^{3,4,5}$, D. Polychroni$^{6,1}$, F. Wyrowski$^{4}$, F. Motte$^7$, D. Elia$^1$, Q. Nguyen-Luong$^8$, P. Didelon$^7$,
\newauthor {\large  M. Pestalozzi$^1$, M. Benedettini$^1$, S. Molinari$^1$, Ph. Andr\'e$^7$, C. Fallscheer$^{9,10}$, A. Gibb$^{11}$, A.~M. di Giorgio$^1$,  T. Hill$^{7,12}$,}
\newauthor {\large  V. K\"onyves$^{7,13}$, A. Marston$^{14}$,  S. Pezzuto$^1$, A. Rivera-Ingraham$^{15,16}$, E. Schisano$^{17}$, N. Schneider$^{18}$, L. Spinoglio$^1$, }
\newauthor {\large D.  Ward-Thompson$^{19}$, and G.~J. White$^{20,21}$}\\ 
$^1$ {\footnotesize Istituto di Astrofisica e Planetologia Spaziali (INAF-IAPS), Via del Fosso del Cavaliere 100, 00133 Roma, Italy} \\ 
$^2$ European Space Research and Technology Centre (ESA-ESTEC), Keplerlaan 1, P.O. Box 299, 2200 AG Noordwijk, The Netherlands\\
$^3$  SKA South Africa, 3rd Floor, The Park, Park Rd, Pinelands, 7405, South Africa \\ 
$^4$ Max-Planck-Institut f\"ur Radioastronomie, Auf dem H\"ugel 69, 53121 Bonn, Germany\\ 
$^5$Hartebeesthoek Radio Astronomy Observatory, PO Box 443, Krugersdorp, 1740, South Africa\\
$^6$ University of Athens, Department of Astrophysics, Astronomy and Mechanics, Faculty of Physics, Panepistimiopolis, 15784 Zografos, Athens, Greece\\
$^7$ Laboratoire AIM Paris-Saclay, CEA/IRFU CNRS/INSU Universit\'e Paris Diderot, 91191 Gif-sur-Yvette, France\\
$^8$ Canadian Institute for Theoretical Astrophysics (CITA), University of Toronto, 60 St. George Street, Toronto, ON, M5S 3H8, Canada\\
$^9$Department of Physics \& Astronomy, University of Victoria, PO Box 355, STN CSC, Victoria, BC, V8W 3P6, Canada\\
$^{10}$National Research Council Canada, 5071 West Saanich Road, Victoria, BC V9E 2E7, Canada\\
$^{11}$ Department of Physics and Astronomy, University of British Columbia, 6224 Agricultural Road, Vancouver, BC, V6T 1Z1 Canada\\
$^{12}$Joint ALMA Observatory, Alonso de Cordova 3107, Vitacura, Santiago, Chile\\
$^{13}$Institut d'Astrophysique Spatiale, UMR8617, CNRS/Universit\'e Paris-Sud 11, 91405 Orsay, France\\
$^{14}$ ESA/ESAC, PO Box 78, 28691 Villanueva de la Canada, Madrid, Spain\\
$^{15}$Universit\'e de Toulouse; UPS-OMP; IRAP;  Toulouse, France\\
$^{16}$CNRS; IRAP; 9 Av. colonel Roche, BP 44346, F-31028 Toulouse cedex 4, France\\
$^{17}$ Infrared Processing and Analysis Center, Institute of Technology, Pasadena, CA, 91125, USA\\
$^{18}$OASU/LAB-UMR5804, CNRS, Universit\'e Bordeaux 1, F-33270 Floirac, France\\
$^{19}$Jeremiah Horrocks Institute, University of Central Lancashire, Preston, Lancashire, PR1 2HE, UK\\
$^{20}$Space Science and Technology Department, STFC Rutherford Appleton Laboratory, Chilton, Didcot, Oxfordshire, OX11 0QX, UK\\
$^{21}$Department of Physics and Astronomy, The Open University, Walton Hall, Milton Keynes, MK7 6AA, UK
}

\begin{document}

   \date{}
\pagerange{\pageref{firstpage}--\pageref{lastpage}} \pubyear{2013}

\maketitle

\label{firstpage}

\begin{abstract}

We present the results of {\em Herschel} HOBYS photometric mapping combined with BIMA observations and additional archival data, and perform an in-depth study of the evolutionary phases of the star-forming clumps in W\,48A and their surroundings. 
Age estimates for the compact sources were derived from bolometric luminosities and envelope masses, which were obtained from the dust continuum emission, and agree within an order of magnitude with age estimates from molecular line and radio data. 
The clumps in W\,48A are linearly aligned by age (east-old to west-young): we find a \hiir , a young stellar object (YSO) with class II methanol maser emission, a YSO with a massive outflow, and finally the NH$_2$D prestellar cores from Pillai et al. 
This remarkable positioning reflects the (star) formation history of the region. We find that it is unlikely that the star formation in the W\,48A molecular cloud was triggered by the \hiir\ and discuss the Aquila supershell expansion as a mayor influence on the evolution of W\,48A. We conclude that the combination of {\em Herschel} continuum data with interferometric molecular line and radio continuum data is important to derive trustworthy age estimates and interpret the origin of large scale structures through kinematic information.
 
\end{abstract}


%

\section{Introduction}

Sequential star formation in the structure of OB associations was noted already in 1960's (e.g., \citealt{blaauw:1964,elmegreen:1977} and references therein). More recently, also on smaller scales stellar age gradients are observed, such as for bright-rimmed clouds (see e.g.,\citealt{white:1997,thompson:2004aa,matsuyanagi:2006,ogura:2007}) and for star formation on the borders of H{\sc ii} regions (e.g.,  \citealt{deharveng:2005,zavagno:2010a}). Evidence for sequential star formation should also be present during the earliest stages of star formation. Molecular clouds containing star-forming objects at various stages of evolution are commonly observed: the {\em Spitzer} study of \citet{qiu:2008} shows the distributions of the various evolutionary classes of young stellar objects (YSOs) within one cloud;  and molecular line studies of infrared dark and high-extinction clouds find collapsing dense infrared-dark cores next to active infrared-bright star-forming ones (see, e.g. \citealt{palau:2010}, \citealt{rygl:2010b,rygl:2013b}). More recently, a few {\em Herschel} HOBYS studies have investigated star formation near OB clusters and bubble-shaped objects in view of triggered star formation (e.g, RCW\,120: \citealt{zavagno:2010}, Rosette: \citealt{schneider:2010,schneider:2012}, \citealt{tremblin:2014}, W5-E: \citealt{deharveng:2012}, W\,3: \citealt{rivera:2013}, NGC 7538: \citealt{fallscheer:2013}). 
It has not been proven yet whether all sequential star formation is the consequence of triggering, and evidence to the contrary has been found for the Rosette region through observations (\citealt{schneider:2012, cambresy:2013}) and simulations (e.g., \citealt{dale:2011,dale:2012}). Determining age gradients across a star-forming region can give insight into the star formation history by which one can assess the possibility of triggering and its size scale. In this paper we analysed the age gradient in W\,48A, a region with a trustworthy distance and various stages of star formation, tapping into new dust continuum and spectral line data as well as archival data. 

W\,48A is part of a group of H{\sc ii} regions (\citealt{onello:1994}) located at a distance of $3.27\pm0.49$\,kpc (maser parallax, \citealt{zhangb:2009}) about 1$\rlap{.}$\degr7 below the Galactic plane. \citet{arthur:2006} describe it as a ``classical" champagne flow plus stellar wind \hiir\ where the photo dissociation region (PDR) is moving with 2.5\,\kms\ into the molecular cloud (\citealt{roshi:2005}). The star-forming region W\,48A  is known to host several different stages of high-mass star formation: an ultra compact (UC) H{\sc ii} region (G\,35.20--1.74, \citealt{wood:1989a}) embedded in a larger, almost 2\arcmin$\times$2\arcmin, cometary part (\citealt{wood:1989a,kurtz:1994,onello:1994, roshi:2005}) extending to the southeast, a methanol maser-emitting YSO (\citealt{minier:2000}) and some recently discovered prestellar cores containing cold and deuterated molecules (\citealt{pillai:2011}). 

W\,48A was observed with the {\em Herschel} Space Observatory (\citealt{pilbratt:2010}) at 70 to 500\,$\mu$m under the {\em Herschel} imaging survey of OB Young Stellar objects programme (HOBYS, \citealt{motte:2010}) as a part of a larger map of the W\,48 molecular cloud complex (see the entire {\em Herschel} maps in \citealt{nguyen:2011}). These observations are a powerful tool to visualise the large scale structure of the cold (160--500\,$\mu$m) and warm (70\,$\mu$m) dust emission.  
{\em Herschel}'s wavelength coverage is ideal to sample the peak of the spectral energy distribution (SED) of cold and warm (10--30\,K) dusty prestellar material and envelopes around YSOs to obtain their temperatures and masses. Since high-mass stars form deeply embedded in their natal cloud, the bulk of their emission is absorbed by the surrounding envelope and is re-emitted in the infrared to mm regime. The infrared luminosity is therefore important for obtaining the bolometric luminosity and determining the spectral type of the embedded star. With the help of a luminosity-mass ($L/M$) diagram (\citealt{saraceno:1996,molinari:2008,elia:2010,hennemann:2010}) one can characterise the evolutionary stage of the YSO through its bolometric luminosity and envelope mass. 

The {\em Herschel} maps were combined with BIMA+IRAM 30m interferometric molecular line imaging to provide a sub-arcsecond resolution view into the star-forming region, down to size scales of $\sim$1000\,AU, and with several archival data sets from the VLA and JCMT. Molecular line and radio continuum data offer an alternative YSO age estimation to that from the $L/M$ diagram. In particular, we used the \amm\,(1,\,1) and (2,\,2) lines to trace the earlier stages of star formation, the CO\,(3--2) line to find evidence of outflows, and the \mecn\,(6--5), $K$=1, 2, 3, 4, transitions and 107\,GHz methanol masers to trace more the hot and dense gas around YSOs. Furthermore we used the molecular line data to obtain kinematical information: methanol masers for the small scale kinematics around the star-forming objects, and \c18o\,(1--0) data for the large scale kinematics.

\section{Observations and data reduction}
\label{sec:obs}
In Table \ref{ta:spec} we give an overview of all the observed continuum wavelengths and molecular transitions, including the observatory, beam sizes, map sizes and the r.m.s image noise. The observations are presented individually, per observatory, in the following subsections. 


\begin{table}
\scriptsize
\begin{flushleft}
\caption{Summary of the observed/archival continuum and spectral line data}
\label{ta:spec}
\tabcolsep=0.11cm
\begin{tabular}{lrllcc}
\hline
Line & \multicolumn{1}{c}{$\nu$/$\lambda$$^{a,b}$} & Observatory & Beam Size$^c$ &Map Size & $\sigma_\mathrm{image}$\\
                                  &  \multicolumn{1}{c}{(GHz/$\mu$m)}       &                      &(\arcsec , \arcsec, \degr )& &(mJy beam$^{-1}$)\\
           \hline  
NH$_3$\,(1, 1) & 23.694 & VLA& 8.8,7.0,42.3&2\arcmin$\times$2\arcmin &1.2 \\
NH$_3$\,(2, 2) & 23.722 &VLA & 8.8,7.0,42.3&2\arcmin$\times$2\arcmin &1.2\\
CH$_3$OH\,(3$_1$--4$_0$) &107.014&BIMA&0.57, 0.24, 45.9&1$\rlap{.}$\arcmin6$\times$1$\rlap{.}$\arcmin6 &250 \\
C$^{18}$O\,(1--0) & 109.782 &B+30m&5.2, 4.1, 1.4&1$\rlap{.}$\arcmin6$\times$1$\rlap{.}$\arcmin6 & 22\\
C$^{18}$O\,(1--0) & 109.782 &IRAM30m&21.8&2$\rlap{.}$\arcmin9$\times$2$\rlap{.}$\arcmin9 &120 \\
$^{13}$CO\,(1--0) &110.198 &BIMA&4.9, 4.0, 2.1&1$\rlap{.}$\arcmin6$\times$1$\rlap{.}$\arcmin6& 45\\ 
CH$_{3}$CN\,(6--5)$^d$&110.383 &BIMA & 4.7, 3.9, 4.9&1$\rlap{.}$\arcmin6$\times$1$\rlap{.}$\arcmin6 & 65\\
CO\,(3--2)& 345.796 &JCMT &14&6$\rlap{.}$\arcmin5$\times$6$\rlap{.}$\arcmin5&12$\times10^3$ \\
continuum & 2800$^b$  &BIMA& 2.3, 1.9, 3.5&1$\rlap{.}$\arcmin6$\times$1$\rlap{.}$\arcmin6 & 3.0\\
continuum & 1250$^b$ &IRAM30m&11 &0$\rlap{.}$\degr3$\times$0$\rlap{.}$\degr3$^e$ & 19\\
continuum & 850$^b$ & JCMT& 22.9$^f$&8$\rlap{.}$\arcmin8$\times$8$\rlap{.}$\arcmin8 &90 \\
continuum & 500$^b$ & {\em Herschel} & 36&2$\rlap{.}$\degr5$\times$2$\rlap{.}$\degr5 & 110\\
continuum & 450$^b$ & JCMT & 17.0$^f$&8$\rlap{.}$\arcmin8$\times$8$\rlap{.}$\arcmin8 & 1.5$\times10^3$\\
continuum & 350$^b$ & {\em Herschel}& 25&2$\rlap{.}$\degr5$\times$2$\rlap{.}$\degr5 & 69\\
continuum & 250$^b$ & {\em Herschel}& 18& 2$\rlap{.}$\degr5$\times$2$\rlap{.}$\degr5 & 110\\
continuum & 160$^b$ & {\em Herschel}& 12&2$\rlap{.}$\degr5$\times$2$\rlap{.}$\degr5 & 34\\
continuum & 70$^b$ & {\em Herschel}& 6&2$\rlap{.}$\degr5$\times$2$\rlap{.}$\degr5 & 16\\
\hline 
\end{tabular}
 
NOTES. Columns are (from left to right): molecular species and transition; frequency or wavelength; observatory; (synthesised) beam size; maps size or field of view (primary beam); r.m.s image noise. $^a$Molecular line frequencies are from the JPL molecular database (\citealt{pickett:1998}). $^b$For the dust continuum the wavelengths (in $\mu$m) are given rather than the frequencies. $^c$ For single-dish maps the beam is assumed to be circular. 10\arcsec corresponds to 0.16\,pc at 3.27\,kpc. 
$^d$Covering the $K$=0, 1, 2, 3 and 4 components. $^e$The 1.2\,mm map is a part of a larger map of 1$\rlap{.}$\degr5$\times$1$\rlap{.}$\degr3, which was not fully covered. $^f$Effective beam sizes as quoted in \citet{difrancesco:2008} rather than the nominal beam sizes.

\end{flushleft}
\end{table}

\subsection{{\em Herschel} observations and data reduction}
\label{sec:obs_herschel}

\begin{figure*}
\includegraphics[angle=-90,width=14cm]{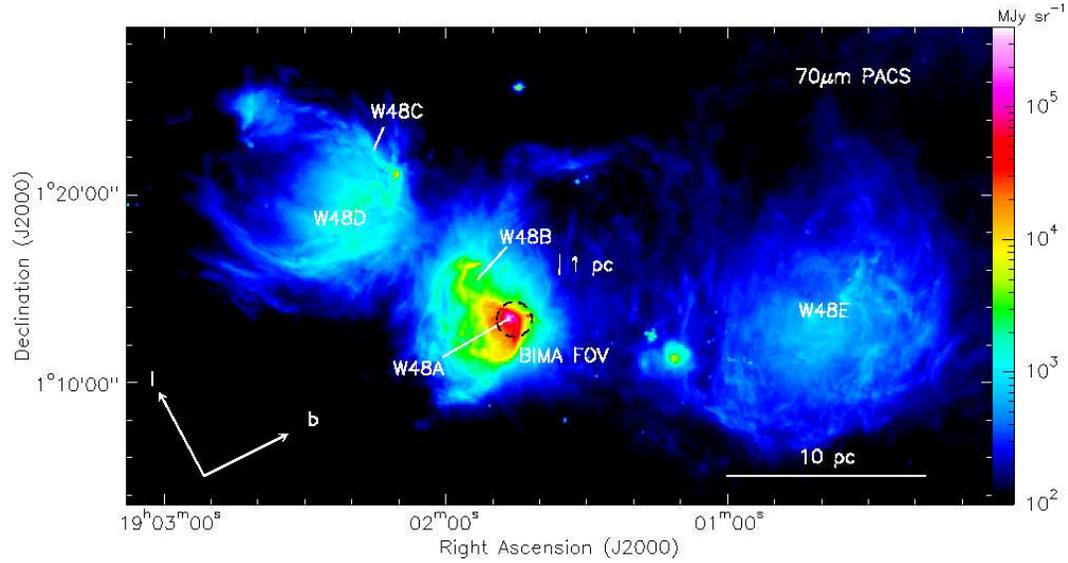}
\caption{{\em Herschel} 70\,$\mu$m map of the W\,48 H{\sc ii} regions. The field of view (FOV) of the BIMA observations is indicated. The orientation of the Galactic coordinates is shown in the bottom left corner.\label{fig:herschel_overview}}
\end{figure*}

\begin{figure*}
\includegraphics[width=4.2cm,angle=-90]{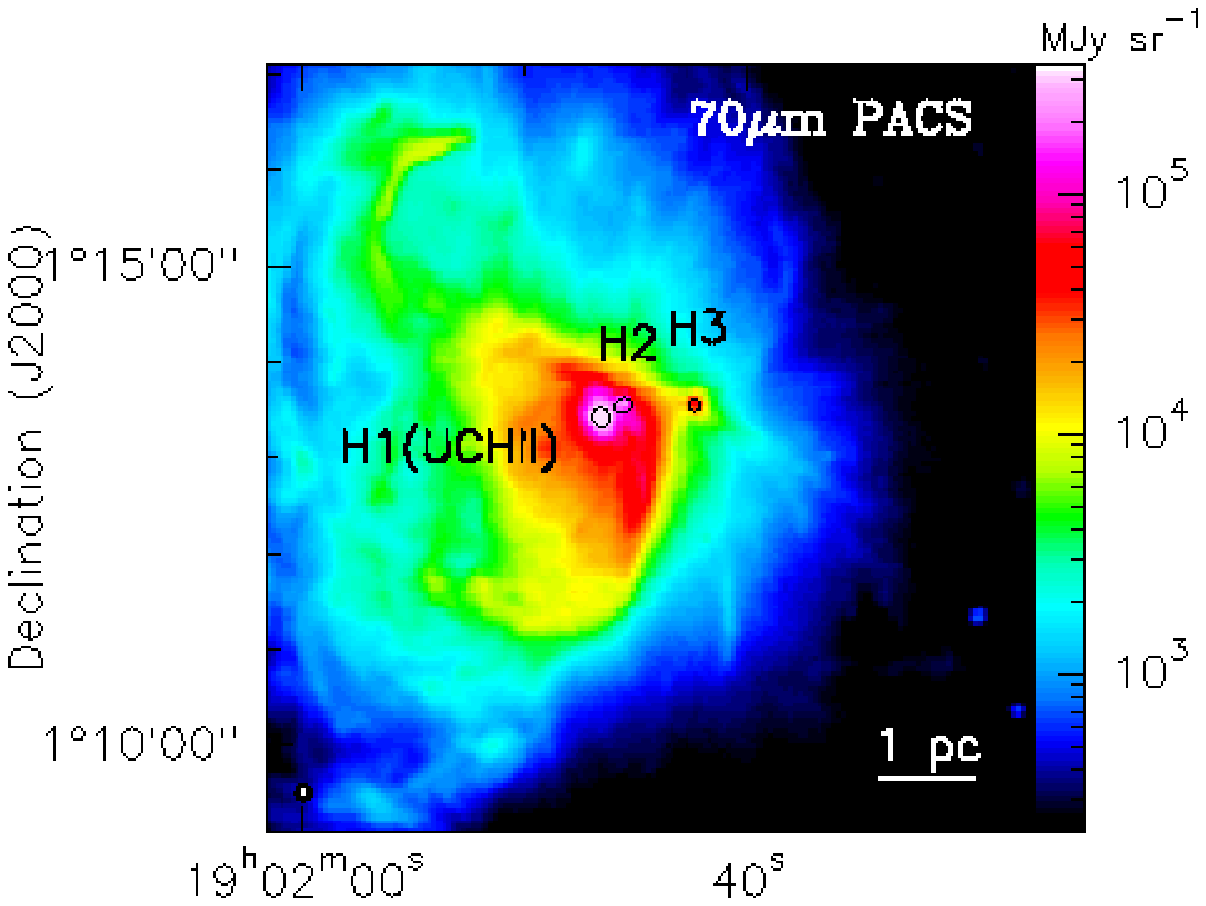}
\includegraphics[width=4.2cm,angle=-90]{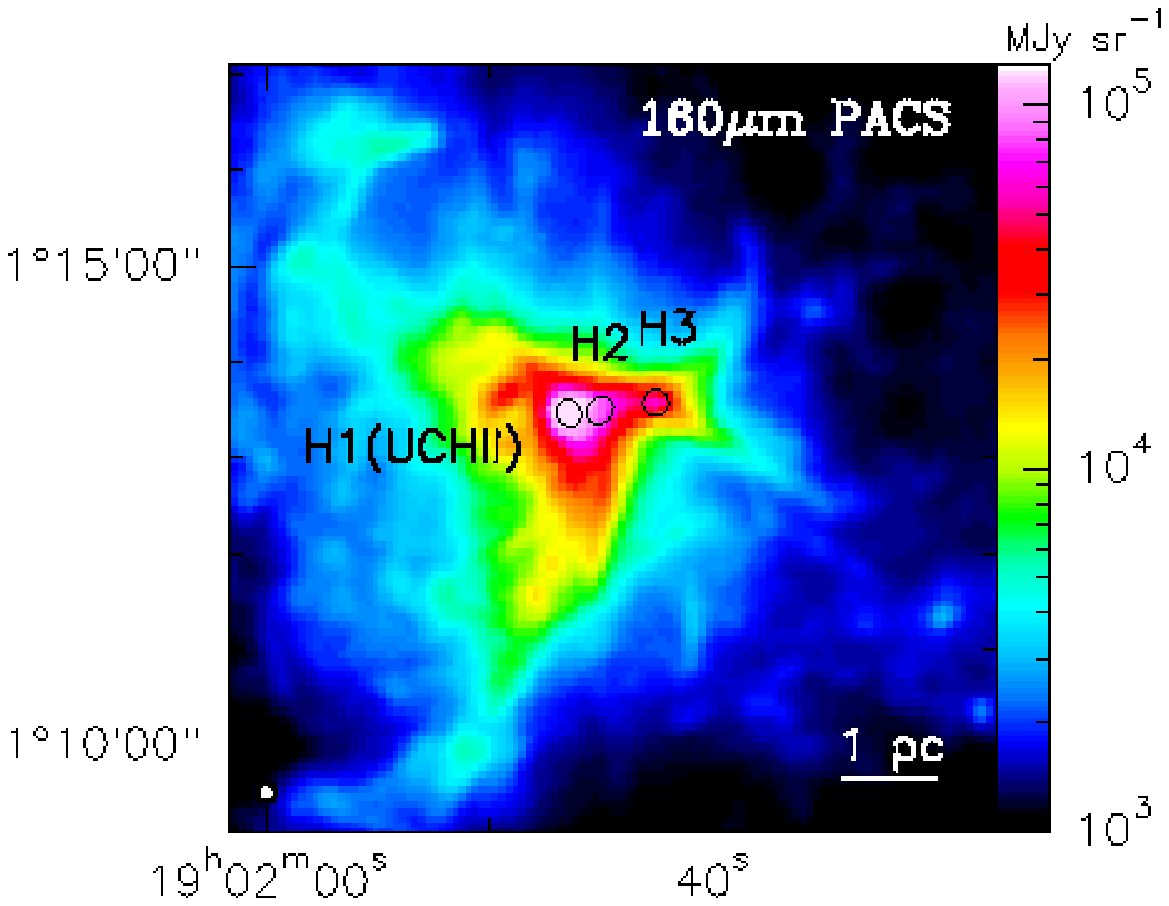}
\includegraphics[width=4.2cm,angle=-90]{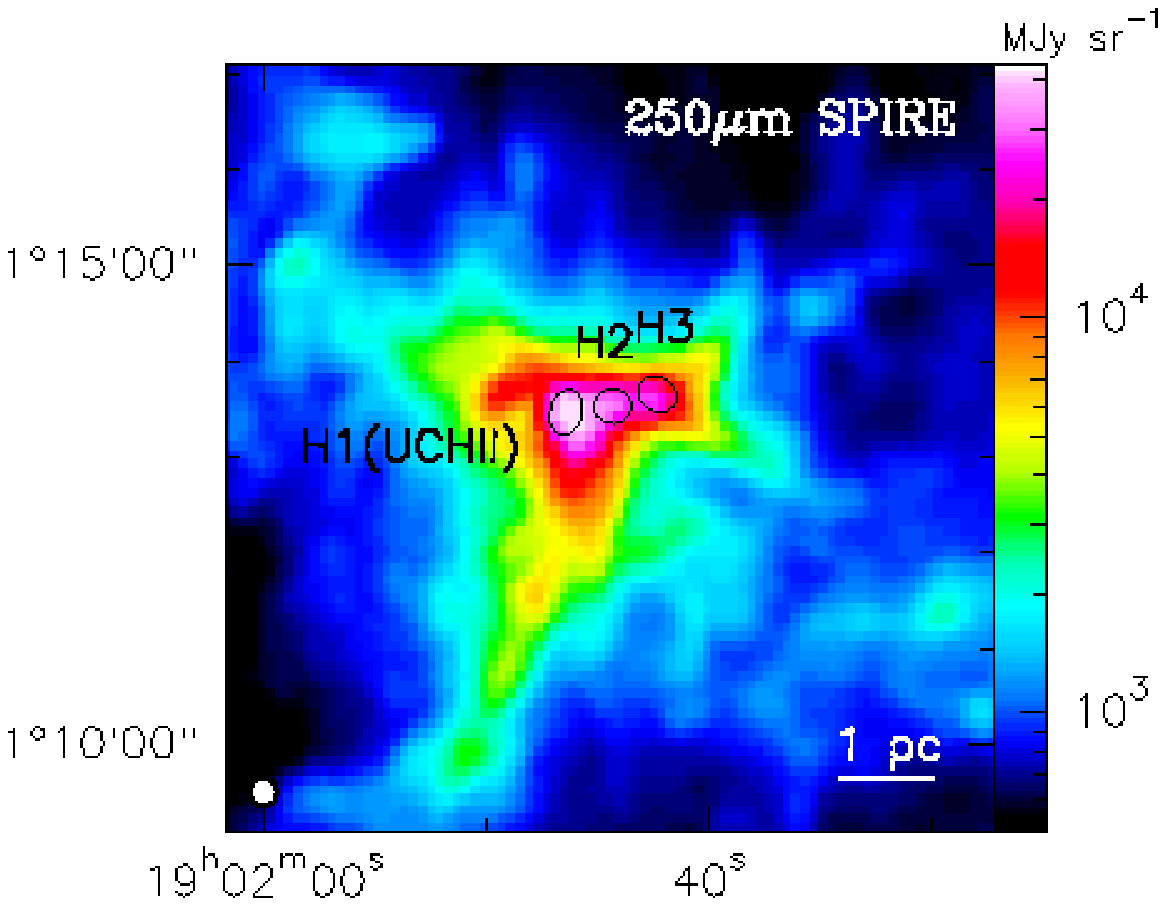}
\includegraphics[width=4.2cm,angle=-90]{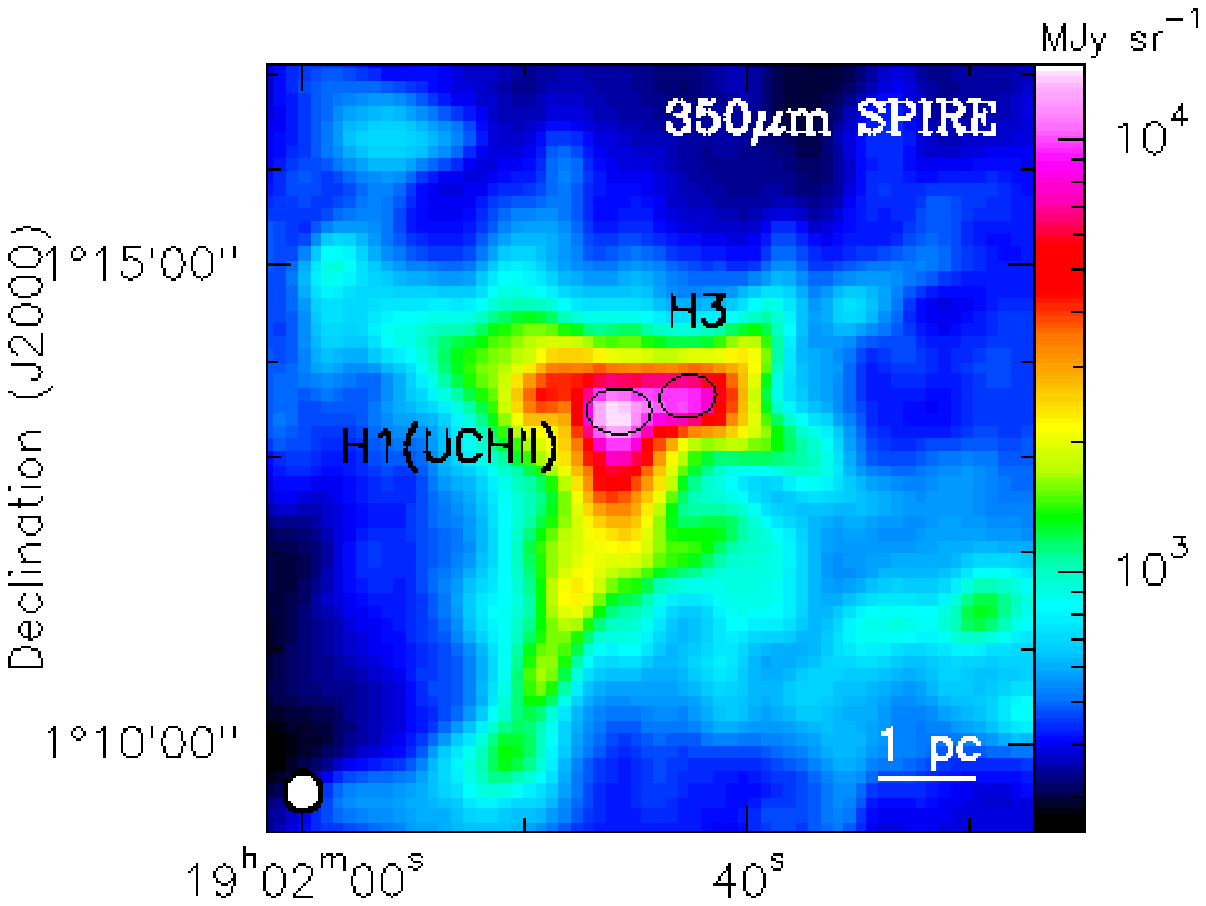}
\includegraphics[width=4.2cm,angle=-90]{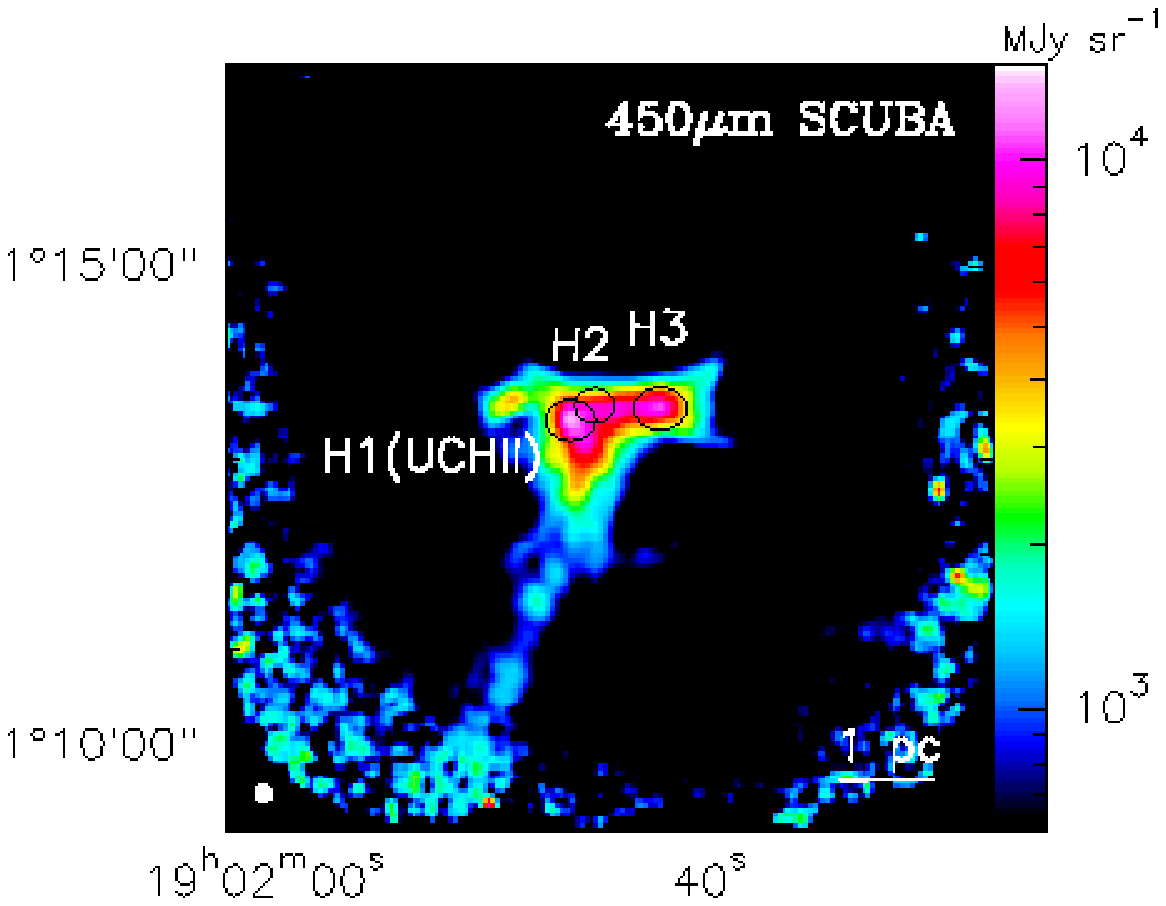}
\includegraphics[width=4.2cm,angle=-90]{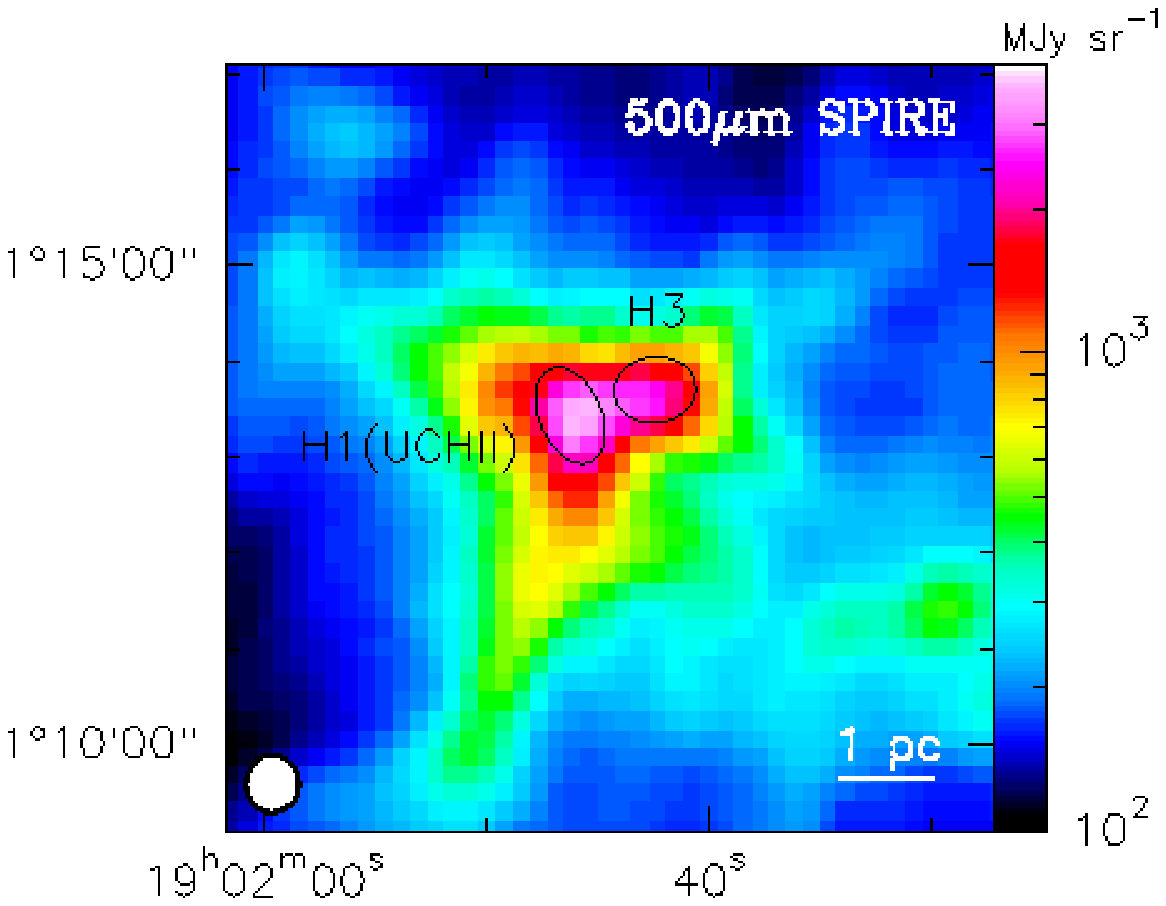}
\includegraphics[width=4.45cm,angle=-90]{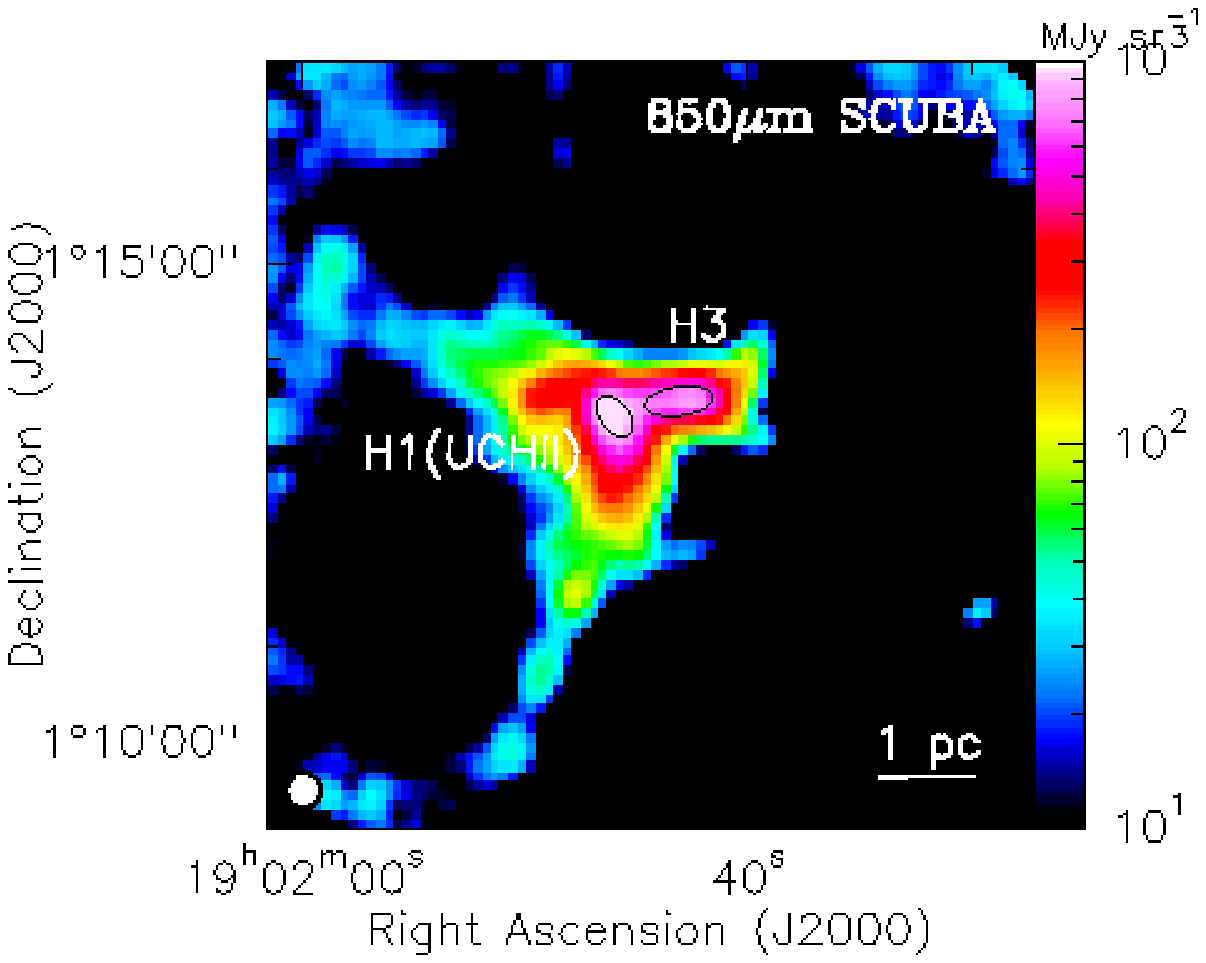}
\includegraphics[width=4.45cm,angle=-90]{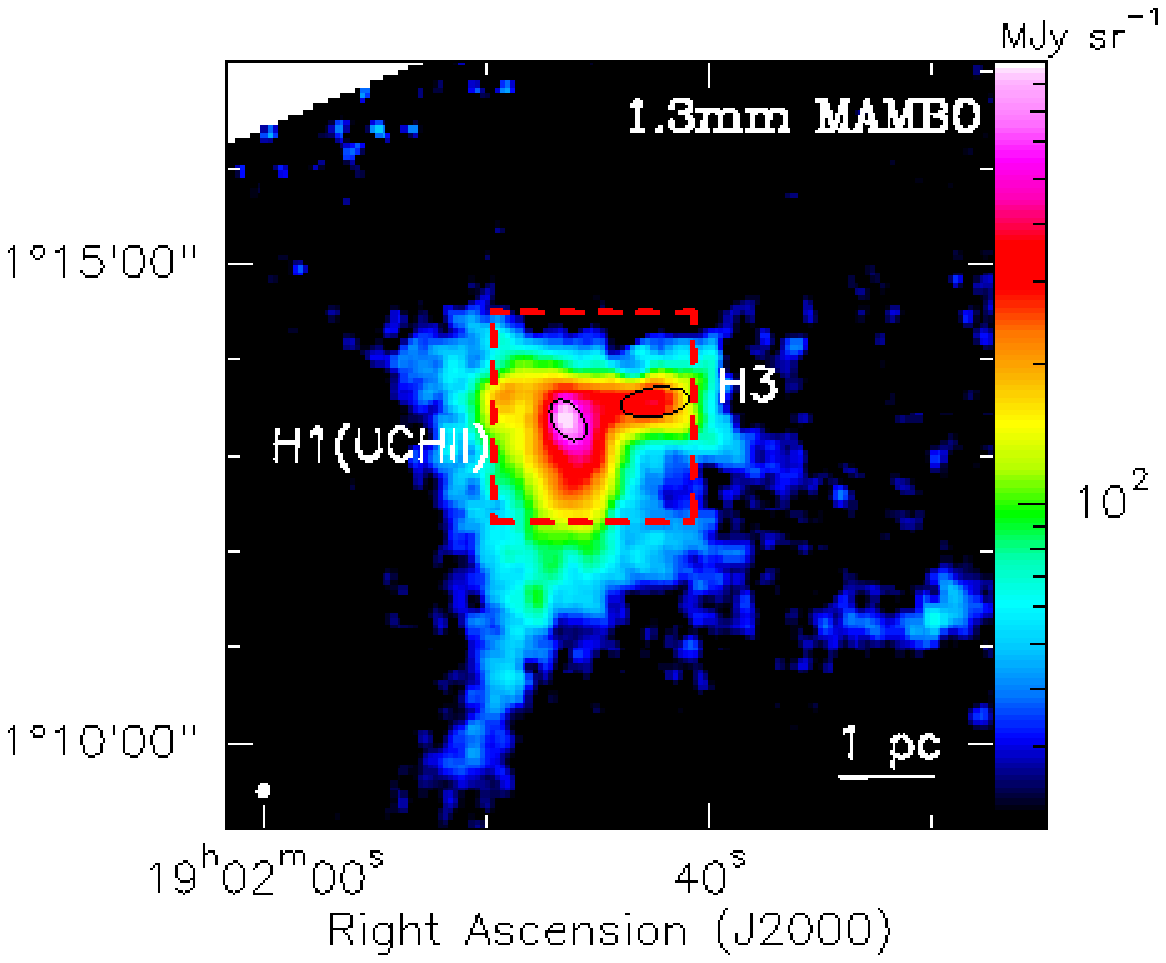}
\includegraphics[width=4.45cm,angle=-90]{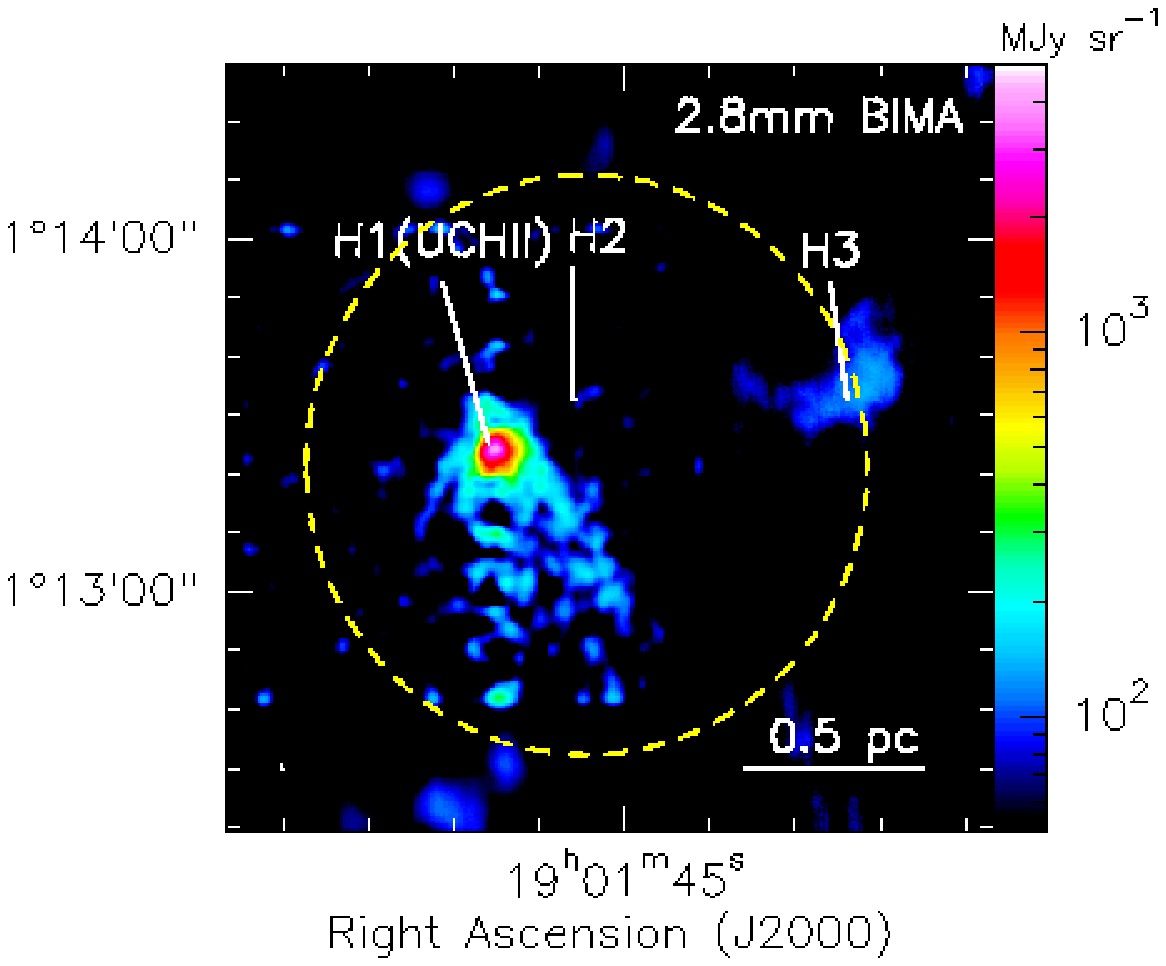}
\caption{\label{fig:continuum_all} PACS, SPIRE, SCUBA, MAMBO and BIMA continuum images of W\,48A. All map units are MJy sr$^{-1}$. Beam sizes are shown in the bottom left corner of each map. The MAMBO map has been smoothed to have a 13\arcsec\ beam. All maps have the same dimension, except for the BIMA image, which is smaller. The red dashed box in the MAMBO map shows the area of the BIMA image. The BIMA field of view is indicated by a yellow dashed circle. For each map, we show with ellipses the  clumps H-1, H-2 and H-3 (see Table \ref{ta:clumps}) as obtained through CuTEx (see Section~\ref{sec:compact_sources}) for that map, except for the BIMA map, where the location of the clumps is given for the reader's convenience and does not refer to a source-detection. }
\end{figure*}

The {\em Herschel} 70, 160, 250, 350, and 500\,$\mu$m observations of W\,48 (ObsIDs: 1342204856 and -7) were carried out in September 2010 using the Photodetector Array Camera and Spectrometer (PACS; \citealt{poglitsch:2010}) and  the Spectral and Photometric Imaging Receiver (SPIRE; \citealt{griffin:2010}) instruments in parallel mode at a scanning speed of 20\arcsec s$^{-1}$. The instruments scanned the W\,48 field in two perpendicular directions, so that each part of the sky was sampled by several bolometer units. The resulting area observed by both PACS and SPIRE covered 2$\rlap{.}$\degr5$\times$2$\rlap{.}$\degr5, including a part of the Galactic plane and reaching slightly beyond the Galactic latitude of --2$\rlap{.}$\degr 1. After exporting the level-0.5 data from HIPE v\,9.0.0 (\citealt{ott:2010}), the data were reduced with the Unimap\footnote{http://w3.uniroma1.it/unimap/} mapmaker v\,5.4.4 (for details see \citealt{traficante:2011} and \citealt{piazzo:2012}). Maps reduced with Unimap are of equivalent quality as those produced with Scanamorphos (\citealt{roussel:2012})  and HIPE, the general map-makers used in the HOBYS programme (e.g., see \citealt{nguyen:2011}). The maps were astrometrically aligned with the 24\,$\mu$m MIPSGAL (\citealt{carey:2009}) data. Absolute (extended) flux calibration of the {\em Herschel} data was done by applying the offsets obtained from the comparison of extended emission from {\em Herschel} data with that from {\em Planck} and {\em IRAS} data, as described in \citet{bernard:2010}, yielding a 20\% flux uncertainty. Bright emission of the  W\,48A \hiir\ caused three pixels to be saturated in the SPIRE 250\,$\mu$m map. To correct for this, W\,48A was observed again (ObsID: 1342244164) on December 4, 2012 with SPIRE in the small map mode (12\arcmin$\times$15\arcmin, cross scan at 30\arcsec/sec scanning speed) and the saturated pixels were replaced. No saturation problems were encountered at the other wavelengths. We estimated the rms image noise (Table~\ref{ta:spec}) for each map from the mean flux of the pixels in a relatively blank region. The conversion from Jy~beam$^{-1}$ to MJy sr$^{-1}$ is Jy~beam$^{-1}\times10^{-6}\times(\frac{\theta}{206265})^{2}\times\frac{\pi}{4\ln{2}}$= MJy sr$^{-1}$, where $\theta$ is the beam FWHM in arcseconds. Figure \ref{fig:herschel_overview} gives an overview of the W\,48 \hiir s (using the naming convention from \citealt{onello:1994}). Maps of W\,48A at all {\em Herschel} wavelengths are shown in Fig.~\ref{fig:continuum_all}. 

\subsection{BIMA observations and data reduction}

Observations were made using the Berkeley Illinois Maryland Association (BIMA) interferometer at 110 GHz (2.8\,mm). All antenna configurations i.e., A, B, C, and D configurations, were used in order to achieve high dynamic range images. The nominal beam sizes at 3\,mm were 0$\rlap{.}$\arcsec 4 for the A configuration, 2\arcsec\ for the B  configuration, 6\arcsec\ for the C  configuration, and 18\arcsec\ for the D  configuration.
The dates of the observations were 2000, December 18, 25, 31 (A configuration); 2003, February 1 (B configuration); 2000, July 14 (C configuration); 2000, May 23 (D configuration). The continuum phase calibrator used was J1751+096.  The pointing centre of the array was 19$^\mathrm{h}$01$^\mathrm{m}$45$\rlap{.}^\mathrm{s}$448 +01\degr13\arcmin21$\rlap{.}$\arcsec49 (J2000). Due to the instability of the atmosphere at $\sim$3\,mm, the observations using the extended A-array configuration were done in fast-switching mode. Figure~\ref{fig:continuum_all} shows the high dynamic range 2.8\,mm continuum map of W\,48A. Our BIMA continuum appears different from that measured at 3.5\,mm with the PdBI by \citet{pillai:2011}, because our BIMA observations are a factor $\sim$4 less sensitive plus our pointing centre was $\sim$50\arcsec east of that used in the observations of \citet{pillai:2011}, decreasing our sensitivity in the western part of W48\,A, which was the objective of their PdBI observations. The pointing accuracy ($\sigma_\mathrm{poi}$) can be estimated following \citet{reid:1988} by 
\begin{equation}
\sigma_\mathrm{poi} = \bigg(\frac{4}{\pi}\bigg)^{0.25} \frac{\theta}{\sqrt{8\ln2}}  \frac{1}{\mathrm{SNR}},
\end{equation}
where $\theta$ is the synthesised beam FWHM in arcseconds and SNR is the maximal flux divided by the rms noise level.  For the BIMA continuum map the pointing accuracy was 0$\rlap{.}$\arcsec02.

Twelve spectral windows were used, targeting the \hiir\ continuum emission as well as line emission from the 107 GHz \meth(3$_{1}$--4$_{0}$, A$^{+}$) masers in the lower sideband and CH$_{3}$CN\,(6--5), $K$=0, 1, 2, 3, 4 levels, C$^{18}$O (1--0) and $^{13}$CO\,(1--0) transitions in the upper sideband.  The channel separations for the lines observed with BIMA are 1.06\,$\mathrm{km~s^{-1}}$ for the \mecn\ lines, 8.50\,$\mathrm{km~s^{-1}}$ for the \thco\ line,  0.27\,$\mathrm{km~s^{-1}}$ for the \c18o\ line and 107\,GHz methanol maser emission. 
For the \c18o\ line, we obtained also a single dish data cube (see Section~\ref{sec:30m}) to have the zero-spacings for complementing the interferometric BIMA data. 

Data reduction was carried out with the \textsc{miriad} package (\citealt{sault:1995}). A phase-delay correction for the lower sideband was found using the continuum phase calibrator. The A-array observations were used to generate a model of the methanol maser. This model was used to self-calibrate the maser observations at all array configurations. The self-calibration solution was then applied to all spectral windows.  The upper sideband observations still had a residual phase error so an additional phase calibration was done on the continuum calibrator in the upper-sideband. The final beam sizes of each map are reported in Table \ref{ta:spec}. 

\subsection{IRAM 30m data}
\label{sec:30m}
\c18o\,(1--0) line observations of W\,48A were carried out in two observing blocks in March 2003 under project number 138-02 at the Institut de Radioastronmie Millim\'etrique (IRAM) 30m telescope. We used the old ABCD receivers in combination with the VESPA backend using a bandwidth of 40\,MHz and a channel spacing of 0.22\,\kms . During the observations, a pointing check was performed approximately every hour on quasar J1749+096, depending on the weather conditions, which were average in block 1 to good in block 2.
The precipitable water vapour (pwv) in block 1 was 6\,mm and system temperatures ranged between 139 to 507\,K. For the second observing block, the pwv was 3--4\,mm with system temperatures of 191 to 292\,K. A focus check was performed on Mars at the beginning of each
observing block. The observed output counts were calibrated to antenna temperatures, $T^\star_{\mathrm{A}}$, using the standard chopper-wheel technique (\citealt{kutner:1981}), and converted to main beam brightness temperatures by multiplying by the ratio of the forward efficiency$^2$ (97\%) and the main beam efficiency\footnote{http://www.iram.es/IRAMES/mainWiki/Iram30mEfficiencies} (54\%) at 110\,GHz. 

The \c18o\ single-dish data were used as the zero-spacing for the \c18o\ interferometric BIMA data. An image cube was made of the BIMA data using maximum entropy deconvolution, excluding a-array data.  Care had to be taken to use an image size smaller than the single-dish image in order to avoid regridding artefacts at the edge of the smaller single-dish image.  The single-dish image was then  regridded to match the BIMA image. A joint maximum entropy deconvolution of the BIMA and IRAM 30m data was done with \textsc{miriad}'s MOSMEM task using rms scaling factors of 1.5 and 2.5 for the BIMA and IRAM 30m data, respectively.

Continuum data were obtained in January 2008 using the Max-Planck Bolometer Array (MAMBO-2, \citealt{kreysa:1998}) under project number 024-07. MAMBO-2 is a 117-pixel array (array size 4\arcmin) that observes at 1.25\,mm (240\,GHz). While these observations covered a large area ($\sim1\rlap{.}$\degr$5\times1\rlap{.}$\degr3), in this paper we use only the part covering the W\,48A \hiir . The MAMBO observations were performed in fast-scanning mode during rough weather conditions (typical opacities at 225\,GHz varied between 0.2 and 0.4). Before each W\,48 map segment a pointing was carried out on the nearby G034.26+00.15 H{\sc ii} region. Calibration was performed using this same H{\sc ii} region, yielding accuracies between 10\% to 30\%. Data reduction was done following the standard pipeline in MOPSIC written by R. Zylka\footnote{http://www.iram.es/IRAMES/mainWiki/CookbookMopsic}. The entire MAMBO map is shown in Fig.~\ref{fig:mambo_allcov}, while the part covering W48\,A is shown in Fig.~\ref{fig:continuum_all}.

\subsection{Archival data}
\subsubsection{JCMT data}

The 450\,$\mu$m and 850\,$\mu$m continuum data from the fundamental catalog of \citet{difrancesco:2008, jenness:2011}\footnote{http://www3.cadc-ccda.hia-iha.nrc-cnrc.gc.ca/community/scubalegacy/} were taken with the Submillimetre Common User Bolometer Array (SCUBA, \citealt{holland:1999}) mounted at the James Clerck Maxwell Telescope (JCMT). The effective beams of the 450\,$\mu$m and 850\,$\mu$m SCUBA data were 17\arcsec and  22$\rlap{.}$\arcsec9 (\citealt{difrancesco:2008}), respectively. Figure \ref{fig:continuum_all} shows the two SCUBA maps of W\,48A.

In addition to the continuum data, we downloaded  raw CO\,(3--2) spectral line data of W\,48A from the JCMT/ACSIS archive\footnote{http://www3.cadc-ccda.hia-iha.nrc-cnrc.gc.ca/jcmt/search/acsis}. These data were obtained on 2008, April 21 with the Heterodyne Array Receiver Program (HARP) array under project name M08AU19 (Lumdsen et al.). HARP is a 16-pixel camera (4$\times$4 array) with a footprint on the sky of 2\arcmin . We reduced the HARP data using the JCMT Starlink ORAC-DR pipeline (\citealt{cavanagh:2008}). The CO data have a channel spacing of 0.42\,$\mathrm{km~s^{-1}}$. The total on-source observation time was $\sim$20 minutes. The observations were carried out in raster-mode using a scan velocity of 7.3\,\arcsec$\mathrm{s^{-1}}$. The final map had 6$\rlap{.}$\arcmin5$\times$6$\rlap{.}$\arcmin5 dimensions, centred on the W\,48A \hiir, at 19$^\mathrm{h}$01$^\mathrm{m}$46$\rlap{.}^\mathrm{s}$7  +01\degr13\arcmin18\arcsec. 

\subsubsection{VLA data}

We collected \amm\,(1,1) and (2,2) data  
from the NRAO's\footnote{The National Radio Astronomy Observatory is a facility of the National Science Foundation operated under cooperative agreement by Associated Universities, Inc.} Very Large Array (VLA) data archive\footnote{https://archive.nrao.edu/archive/advquery.jsp}. The ammonia lines were observed under program AM652 (Minier et al.) on 2000, August 3 with the VLA in D-configuration. 
W\,48 was observed in dual polarisation using two 64 channel intermediate frequency (IF) bands, with the first IF centred on \amm\,(1,1) and the other on \amm\,(2,2). The channel width of the two maps was $\sim$0.62$\,\mathrm{km~s^{-1}}$, enough to resolve the hyperfine structure of the \amm\,(1,1) line. The hyperfine structure of \amm(2,2) is wider (see \citealt{barrett:1977}) than that of \amm(1,1): the inner satellites were located at the border of the observed band and could not be detected. The total on-source integration time was $\sim$50 minutes. The VLA ammonia maps were obtained in one pointing centred on 19$^\mathrm{h}$01$^\mathrm{m}$45$\rlap{.}$$^\mathrm{s}$5 +01\degr13\arcmin28\arcsec, thus covering the entire W\,48A region in the primary beam (FWHM $\sim$2\arcmin, Table \ref{ta:spec}). The data reduction was done following the standard AIPS (\citealt{greisen:2003}) and {\sc miriad} (\citealt{sault:1995}) routines.

\section{Results and analysis of compact objects}

\subsection{Cold dust emission from {\em Herschel} to IRAM 30m}
\subsubsection{Source extraction and SED fitting}
\label{sec:compact_sources}

W\,48A's compact source detection and extraction was performed with the Curvature Threshold Extractor (CuTEx, \citealt{molinari:2011}) in each of the five Herschel bands, the SCUBA 450 and 850\,$\mu$m map, and the MAMBO 1.3\,mm map. CuTEx uses the second derivative of the emission map to obtain source locations, and then derives the source parameters, such as their size, peak flux etc., through 2-dimensional Gaussian fitting (see \citealt{molinari:2011} for more details). We used a curvature threshold of 1.5, and required the source detection to have signal-to-noise level higher than 3.0. Sources across the eight bands were associated according to their positions, requiring the positional distance to be within the radius of the FWHM beam size at the longer wavelength (see \citealt{elia:2010} for more details). The deconvolved sizes of the compact sources were found to increase with wavelength, which is commonly found in previous {\em Herschel} studies (see e.g., \citealt{nguyen:2011,giannini:2012}) for sources embedded within a large-scale emitting structure such as a low density envelope. In Table \ref{ta:spec} one can see that the beam size of the Herschel bands increases with wavelength. Therefore, at longer wavelengths the star-forming clump is more confused with its lower density envelope, which causes its size and intensity to be overestimated. One can correct for this by assuming a density profile for the dust envelope. 

Following  the method introduced by \citet{motte:2010} we assumed a density profile of $n\propto r^{-2}$, and scaled the intensity of the SPIRE 250, 350, 500\,$\mu$m and SCUBA 850\,$\mu$m sources by the ratio of the deconvolved source size at 160\,$\mu$m and that at 250, 350, 500 and 850\,$\mu$m, respectively (see \citealt{nguyen:2011} for a detailed explanation of the flux scaling). In this scaling we assume that the emission between 160 and 850\,$\mu$m is optically thin and emanates from the same volume of gas. From previous SED analysis of {\em Herschel} sources (e.g., \citealt{giannini:2012}) these assumptions appear valid. The 450\,$\mu$m point source fluxes were not flux-scaled since the sources were unresolved within the 17\arcsec\ beam.

We found three compact sources in the W\,48A region, clumps H-1, H-2 and H-3. Clump H-1 was detected in all wavelengths, from 70\,$\mu$m down to 2.8\,mm (and to cm wavelengths), clump H-2 was not detectable longward of 450\,$\mu$m, and clump H-3 was detected down to 1.3\,mm. Using the convention of \citet{henning:2008},  objects with radii$\geq$0.1\,pc are called clumps, while objects with radii$<$0.1\,pc are called cores (see Table~\ref{ta:clumps} for the radii). CuTEx results of these clumps at each wavelength are shown by ellipses in Fig.~\ref{fig:continuum_all}. The clumps were found to be well separated from each other, and did not suffer from blending at the detected wavelengths. The only exception is clump H-2 at 350 and 500\,$\mu$m: this clump was not detected at these wavelengths and it is possible that its emission was confused with that of clump H-1. Clump H-1 coincides with the W\,48A \hiir\ (G35.20--1.74), and is the only one with a previous $IRAS$ detection (\citealt{wood:1989a}). Clump H-2 is spatially coincident with the location of several maser species: water masers (\citealt{hofner:1996}), hydroxyl masers (\citealt{caswell:2001}), and methanol masers (e.g., \citealt{minier:2000,valtts:2002}). Clump H-3 coincides with the 850\,$\mu$m continuum source W\,48W (\citealt{curran:2004}) and a water maser detection (\citealt{hofner:1996}). 

\begin{table*}
\begin{minipage}{180cm}
\begin{flushleft}
\caption{Cold dust continuum properties of W\,48A clumps and cores\label{ta:clumps}}
\begin{tabular}{l c c c c c c c c c c  c c}
\hline
Object & R.A. (J2000)&Dec. (J2000) & $R_\mathrm{160}^a$& $R_\mathrm{CH_3CN}$&$M_\mathrm{env}$ & $T_\mathrm{dust}$ & $\beta$ &$L_\mathrm{bol}$ & Age & Age range\\
 & (h:m:s) & (\degr:\arcmin:\arcsec) &pc&pc&($M_\odot$)& (K) & &($L_\odot$)&($10^5$\,yr) &($10^5$\,yr) \\ 
\hline
clump H-1 &19:01:46.6   & 01:13:25&0.10&& 720$\pm$250 & 37$\pm$4&1.5$\pm$0.1 & 77200$\pm$15000 &7.5 & 2--13\\ 
clump H-2 &19:01:45.7 & 01:13:33&0.11&&240$\pm$60 & 27$\pm$4 & 2.0$^b$&11200$\pm$500&15.1 & 2.4--12\\ 
clump H-3 &19:01:42.4 & 01:13:33&0.10&&1100$\pm$600 & 21$\pm$4 &1.7$\pm$0.2&4000$\pm$1500&1.2 & 0.9--1.5\\ 
core H-2a &19:01:45.5 & 01:13:33& &0.05&170$\pm$30&27$\pm$9&2.0$^b$&8000$\pm$1000&8.0 & 6.8--9.5\\
core H-3b & 19:01:42.3 & 01:13:33&&0.08&420$\pm$200&19$\pm$4&1.9$\pm$0.3&2700$\pm$600&1.5& 1.0--2.0\\

\hline
\end{tabular}
 
NOTES. Columns are (from left to right): object name; positions of clumps (continuum) and cores (\mecn); clump radius at 160\,$\mu$m; radius of \mecn\ core; \\
envelope mass; dust temperature; grain emissivity parameter; bolometric luminosity;  $L/M$-based average age estimation and the range of ages from the two \\
nearest star formation tracks. $^a$ Radius is the half of the full width at half maximum of the Gaussian fit performed with CuTEx on the 160$\,\mu$m map and \\
deconvolved from the beam at 160$\,\mu$m, 12\arcsec . $^b$ For clump H-2 and core H-2a the emissivity index $\beta$ was fixed at 2.0 because the SEDs had no millimetric\\
data points.\\
\end{flushleft}
\end{minipage}
\end{table*}

\begin{figure}
\includegraphics[angle=-90,width=8.5cm]{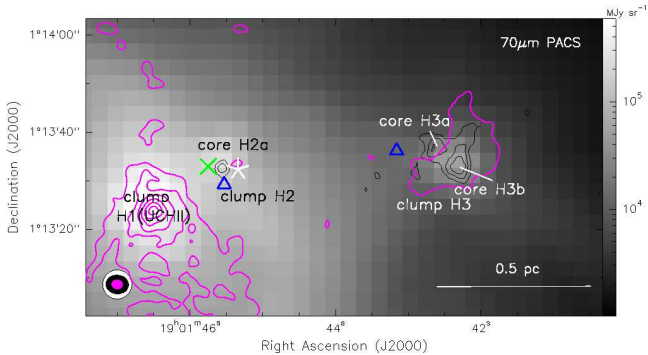}
\caption{\label{fig:maser_ch3cn} 70\,$\mu$m PACS map overlaid with black contours for the total intensity \mecn\ map integrated over the $K$=0 and $K$=1 emission lines showing the locations of the \mecn\ cores, and magenta contours for the 2.8\,mm BIMA continuum. The beam sizes of the 70\,$\mu$m (white), \mecn\ (black), and the 2.8\,mm (magenta) maps are indicated in the bottom left corner. The \mecn\ contours start at 3$\sigma$ (0.06 Jy\,beam$^{-1}$\,km\,s$^{-1}$) and range up to 1.2 Jy\,beam$^{-1}$\,km\,s$^{-1}$ with increments of 0.02 Jy\,beam$^{-1}$\,km\,s$^{-1}$. The 2.8\,mm contours start at 3$\sigma$ (9mJy) with steps by a factor 3. Marked are the maser species observed: hydroxyl masers (green cross, \citealt{caswell:2001}), methanol (white star, this work) and water masers (blue triangle, \citealt{hofner:1996}). The masers around core H-2a have been offset from the core centre artificially. The locations of clumps and cores are indicated on the map.}
\end{figure}

Table \ref{ta:clumps} gives the clump (and core) positions. It is known that clumps can fragment into smaller components (cores) at higher angular resolutions (\citealt{zhang:2009,bontemps:2010b,hennemann:2010,wang:2011,palau:2013}). Using our interferometric \mecn\ data (see Section~\ref{sec:mecn}), we found that clump H-2, contains a more compact core, which we name core H-2a, and that clump H-3 fragments into two \mecn\ cores, H-3a and H-3b (see Fig.~\ref{fig:maser_ch3cn}). Using the same density profile, $n\propto r^{-2}$, we rescaled the continuum fluxes to the sizes of \mecn\ cores (see Table \ref{ta:clumps}). The fluxes of clump H-3 were rescaled to the size of core H-3b rather than core H-3a, since the location core H-3b coincides with the centre of clump H-3, while H-3a does not (Fig.~\ref{fig:maser_ch3cn}).  Moreover, toward core H-3b we detected extended 2.8\,mm continuum emission (BIMA), while non was found toward core H-3a. Also in the PdBI data of \citet{pillai:2011} core H-3a does not coincide with any compact continuum emission, while core H-3b coincides with their PdBI mm1 core. Due to the complete absence of continuum emission toward core H-3a we do {\em not} consider it as a star-forming object, and we attribute all the continuum emission to H-3b.

\begin{figure}
\centering
\includegraphics[width=16cm,angle=-90]{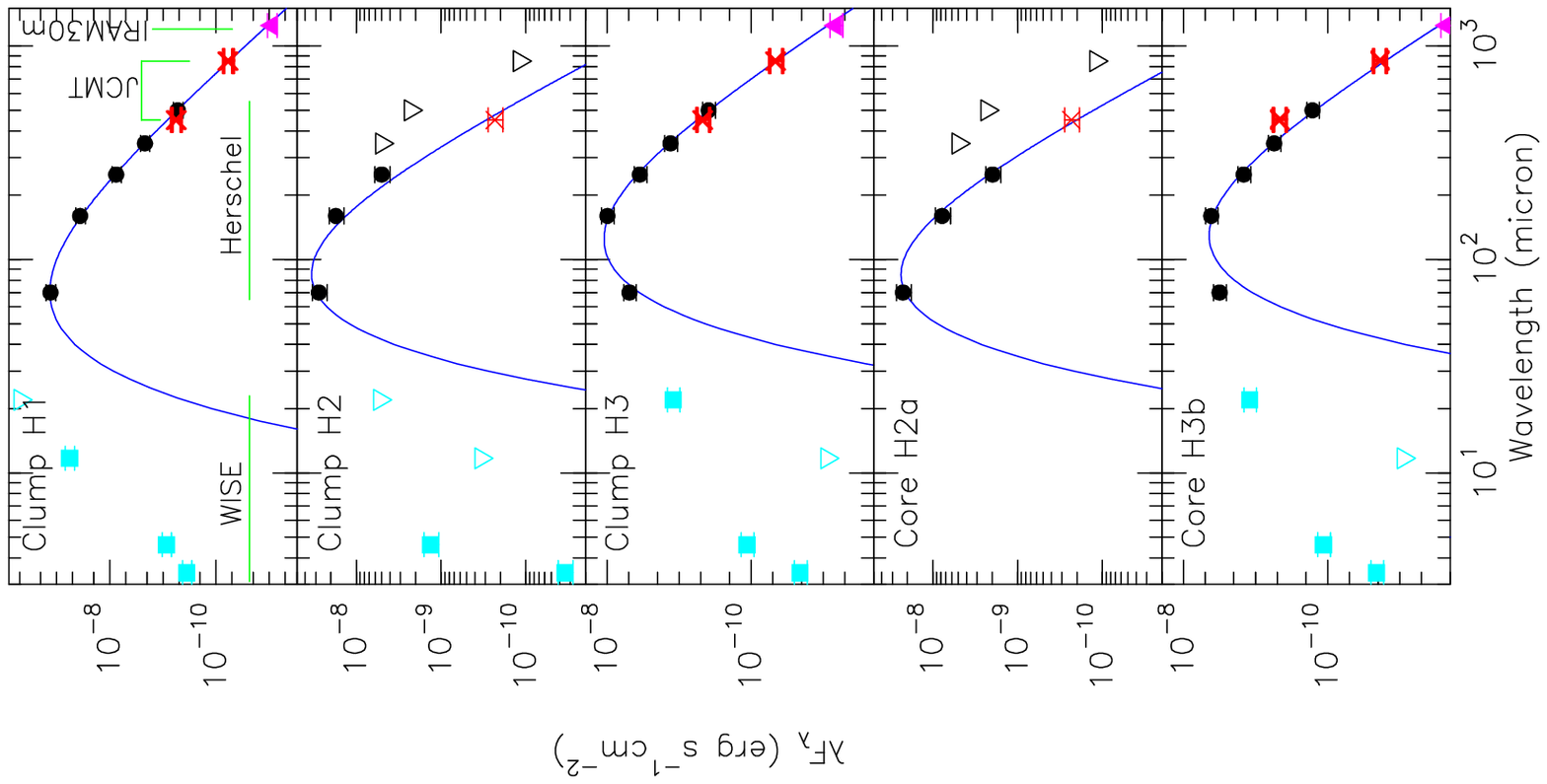}
\caption{\label{fig:seds}Spectral energy distributions of clumps H-1, H-2 and H-3 and cores H-2a and H-3b. Downward facing triangles indicate 3$\sigma$ upper limits. The grey-body fit to the cold dust component is shown by a blue line. The estimated masses, temperatures, dust emissivity, and bolometric luminosities are listed in Table \ref{ta:clumps}. Data points from the various observatories are colour coded: cyan squares for WISE, black filled dots for {\em Herschel}, red crosses for JCMT, and violet filled triangles for IRAM 30m}
\end{figure}

To obtain masses and temperatures we fitted a single-temperature modified black body function to all the continuum fluxes assuming flux uncertainties of 20\%. The grain emissivity parameter $\beta$ was allowed to vary between 1 and 2.5 (\citealt{sadavoy:2013}) for clumps/cores with ancillary millimetric data and kept constant at 2.0 otherwise. Errors in temperature, mass, $\beta$ and luminosity were estimated by varying the continuum fluxes ($\pm$20\%) and estimating how much these quantities changed.
The clump and core SEDs are shown in Fig.\,\ref{fig:seds} with the best modified black body fit, while the resulting masses, temperatures and $\beta$'s are given in Table \ref{ta:clumps}. From the fit to the SED of clumps H-1 and H-3 (Fig.~\ref{fig:seds}) one can see the that the scaled {\em Herschel} and JCMT (850\,$\mu$m) fluxes match well with unscaled the IRAM 30m data points justifying the flux scaling on the 160\,$\mu$m size. Thanks to the SCUBA 450$\,\mu$m data we could confine the clump H-2/core H-2a SED and obtain relatively good estimates of masses, temperatures and luminosities. However, because this source was lacking millimetre detections we kept $\beta=2.0$.  We find that clumps H-1 and H-2 are warmer and less massive than clump H-3. 
Rescaling the clump fluxes with the core sizes did not significantly affect the temperatures, but strongly reduced the envelope masses. 

\subsubsection{Millimetre emission and free-free contamination}
Close to H{\sc ii} regions, free-free emission will be present in addition to thermal dust emission. Free-free emission dominates at centimetre frequencies, and contributes to the thermal emission at millimetre and sub-millimetre wavelengths. Using the radio free-free emission of W\,48A \hiir\ reported by \citet{roshi:2005}, we extrapolated the free-free contribution at 850$\,\mu$m (10\% - 1.76\,Jy) and 1.2\,mm (32\% - 1.80\,Jy) and corrected the clump H-1 fluxes at these wavelengths, which were used for SED fitting, accordingly. In the 2.8\,mm map (Fig.~\ref{fig:continuum_all}), the \hiir\ is clearly visibly as a bright source with some weak extended emission flowing southwards, which is thought to be associated with the compact centimetre component identified by \citet{kurtz:2002}. Toward clump H-2, we found very weak 2.8\,mm emission, at about a 2.5$\sigma$ level given the image noise of 3 mJy\,beam$^{-1}$, not significant enough for our selection criteria.  The millimetre emission toward clump H-3 is a bit stronger (about 3--4$\sigma$) and its shape correlates well the cold {\em Herschel} dust continuum emission (Fig.~\ref{fig:continuum_all}), suggesting that it is real. The extrapolation of the modified black body fit performed on the extended emission in the {\em Herschel} bands would predict about 16\,mJy per 11$\rlap{.}$\arcsec 5 pixel, while the BIMA map contains about 20\,mJy per 11$\rlap{.}$\arcsec 5 pixel in that area. We conclude that most of the extended 2.8\,mm flux around clump H-3 originates in thermal dust emission. 

\subsubsection{Bolometric luminosities and age estimations}

Objects with embedded YSOs will emit strongly not only in the wavelengths $\geq$70\,$\mu$m, but also in the mid and near infrared. 
To obtain a realistic estimate of the bolometric luminosity, we integrated the fluxes between 3\,$\mu$m and 1.3\,mm, using our obtained flux measurements complemented with, when available, the mid-infrared (3.4--22\,$\mu$m) fluxes from the WISE catalog (\citealt{wrightwise:2010}). All data points used to obtain the bolometric luminosity are plotted in Fig.\,\ref{fig:seds} and the resulting luminosity $L_\mathrm{bol}$ is given in Table \ref{ta:clumps}.

For clump H-1, which contains the \hiir , we estimated the stellar type based on the bolometric luminosity, under the assumption that the luminosity is dominated by the most massive object, using the Zero-Age Main-Sequence (ZAMS) column of Table 1 of \citet{panagia:1973}. With respect to the previous, lower resolution $IRAS$ observations (\citealt{wood:1989a}), we find a lower bolometric luminosity and a slightly later spectral type (O7.5 with respect to the previous O6) due to a lower level of confusion. Using the more recent O star parameters of \citet{martins:2005} the spectral type becomes O8. Our spectral type result, O7.5 or O8, reconciles the infrared-based spectral type with the one determined from radio properties -- O7.5 (\citealt{wood:1989a}) and O8 (\citealt{roshi:2005}). 

\begin{figure}
\centering
\includegraphics[width=8.5cm]{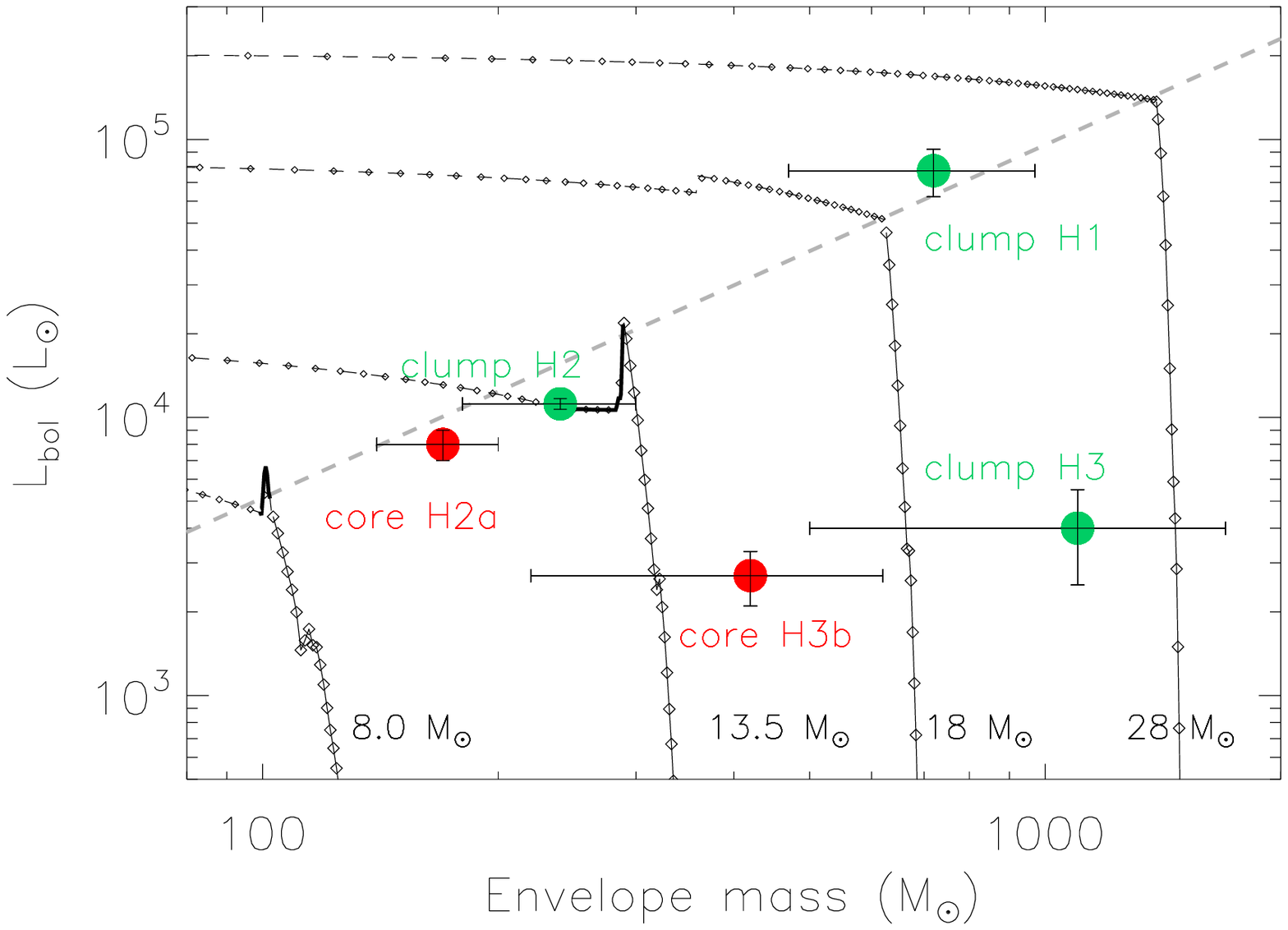}
\caption{\label{fig:lmdiag} Bolometric luminosity versus envelope mass diagram for the sources in Table~\ref{ta:clumps}. Clump H-1, H-2, and H-3 are shown in green and cores H-2a, and H-3b are shown in red. Stellar evolutionary tracks from \citet{molinari:2008} are plotted with the final stellar mass near each track. Open diamonds mark the envelope accretion stage (pre main sequence objects): the symbols are in steps of $10^4$ yr. Open dots mark the envelope dispersed stage (ZAMS star already formed): symbols are in steps of $10^5$yr.  The dashed grey line shows the log-log fit to the IR-primary sources (objects that could be fitted with an embedded ZAMS model and that had the most massive envelope) from \citet{molinari:2008} representing the ZAMS line.}
\end{figure}

Bolometric luminosities and envelope masses are particularly important quantities, because they allow us to assign an age estimation to the object that is forming in the clump or core following the evolutionary tracks from \citet{saraceno:1996} for low-mass stars and \citet{molinari:2008} for high-mass stars. In the $L/M$ diagram, stellar evolution is represented in two main phases: the stellar accretion phase and the envelope clean-up phase. The first phase is when the forming star is accreting matter from the envelope, which increases its luminosity. Mass accretion stops when the star begins to burn hydrogen burning, i.e. when it arrives on the ZAMS.  During the second phase the protostellar envelope is dispersing, while the luminosity stays roughly constant (within the same order of magnitude). 

The stellar evolution tracks in $L/M$ diagram assume the formation of single objects. A rough idea of the impact of multiplicity can be made invoking the Salpeter's initial mass function: 
\begin{equation}
N=k_1 M^{-1.35},
\end{equation}
where $N$ is the number of objects, and $M$ is stellar mass. It shows that most of the mass is in lower-mass objects. When the luminosity is determined by the stellar luminosity, it can be related to the mass by $L\propto M^{3.5}$ and it will dominated by the more massive objects. Hence the envelope mass will be much more affected by the multiplicity than the luminosity. 

Figure \ref{fig:lmdiag} shows the location of clumps H-1, H-2, H-3, and cores H-2a and H-3b on the $L/M$ diagram. We estimated the age range from the two stellar tracks that surround the clump or core, for example for clump H-1 we used the 18 and 28\,$M_\odot$ stellar tracks. The obtained age ranges and average ages are listed in Table \ref{ta:clumps}. The difference between using the masses and luminosities of cores rather than the clumps is that cores are less massive and slightly less luminous. This puts the cores on less massive evolutionary tracks which are evolving slower, and hence the ages become larger. Since the $L/M$ diagram assumes the objects to be single, the core ages should be more accurate than those of the clumps and in the rest of the paper we will use the ages of cores H-2a and H-3b, rather than those of clumps H-2 and H-3, 
For clump H-3 to core H-3b the evolutionary stage did not change very much: both objects are in the envelope accretion phase and their ages are similar. The age difference between clump H-2 and core H-2a in the $L/M$ diagram is a little larger due to the more complex behaviour of the stellar tracks in the $L/M$ diagram near the ZAMS line. In the following sections, we present molecular line data to find further relative and absolute age estimates to compare with the $L/M$ based ages.

\subsection{Class II methanol masers}
\label{sec:maser}

We detected the class II $3_1-4_0, A^+$ methanol maser transition at 107\,GHz toward clump H-2/core H-2a. This is the first interferometric observation of this maser transition in W\,48A delivering  sub-arcsecond images of the maser spots and their distribution, as previous 107\,GHz detections were single-dish only (\citealt{valtts:1995}). The distribution of 107\,GHz methanol masers spots as well as their correlation with the 6.7\,GHz and 12.2\,GHz methanol masers (\citealt{minier:2000}) is discussed in Appendix~\ref{sec:maser_app}. The class II methanol maser detection toward clump H-2 is important, as class II methanol masers are thought to be exclusively associated to high-mass star formation (\citealt{menten:1991,minier:2003}), which, despite some recent doubts, has been confirmed again (\citealt{breen:2013}). Class II methanol masers are excited before and during the onset of the H{\sc ii} region, and hence to have younger or similar ages i.e., $\lesssim10^5$\, yr (\citealt{walsh:1998,codella:2000,breen:2010}). 

\subsection{Methyl cyanide (\mecn) emission}
\label{sec:mecn}

\begin{table}
\footnotesize
\caption{Derived properties of the CH$_{3}$CN cores.}
\label{tab:ch3cn}
\tabcolsep=0.09cm
\begin{tabular}{ccccccccc}
\hline
 Core &  R.A. & Dec. & R & V$_\mathrm{LSR}$     & $\Delta v_\mathrm{obs}$ &T$_\mathrm{rot}$  & $N_\mathrm{CH_3CN}$\\
          & (h:m:s)   & (\degr : \arcmin : \arcsec) & (pc) &(km~s$^{-1}$) &(km~s$^{-1}$) &  (K)               &   (10$^{12}$cm$^{-2}$)\\
          \hline
 H-2a       &19:01:45.55 & 01:13:33.0 & 0.05&42.7$\pm$0.2  & 1.9$\pm$0.4 &  61$\pm$20  &  8$\pm$3\\
 H-3a       & 19:01:42.65 & 01:13:37.5 &0.06&42.1$\pm$0.1  & 2.2$\pm$0.4 & 44$\pm$4 &  9$\pm$6\\
 H-3b      &  19:01:42.31 & 01:13:33.5  &0.08&42.0$\pm$0.2 & 2.3$\pm$0.5& 41$\pm$6 &  8$\pm$2\\
 \hline
\end{tabular}  

NOTES. Columns are (from left to right): core name; positions (J2000); radius of \mecn\ core; \mecn\ LSR velocity; \mecn\ line width of $K=0$ level; \mecn\ rotational temperature; \mecn\ column density\\
\end{table}

\begin{figure*}
\includegraphics[angle=-90,width=5.81cm]{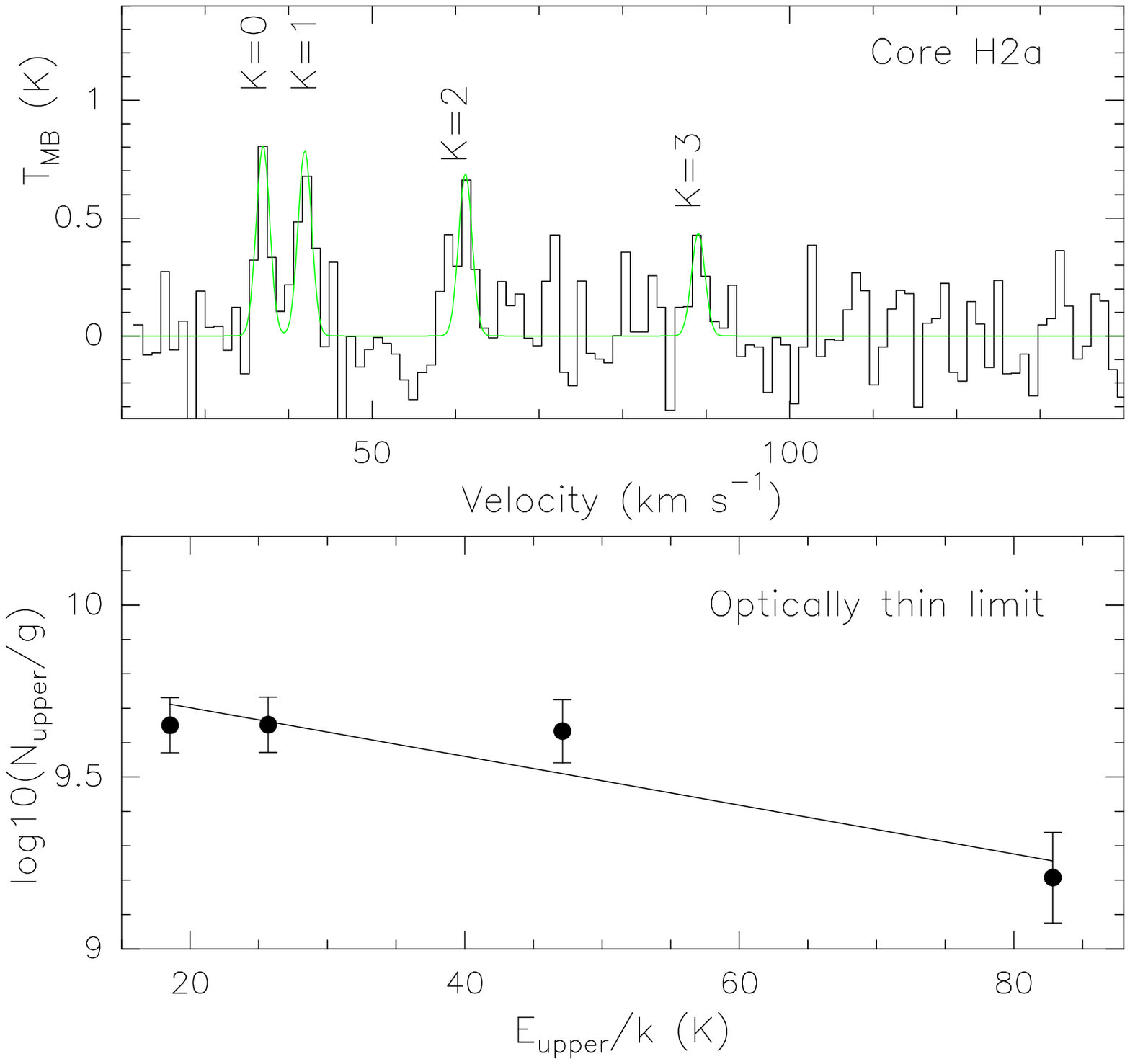}
\includegraphics[angle=-90,width=5.5cm]{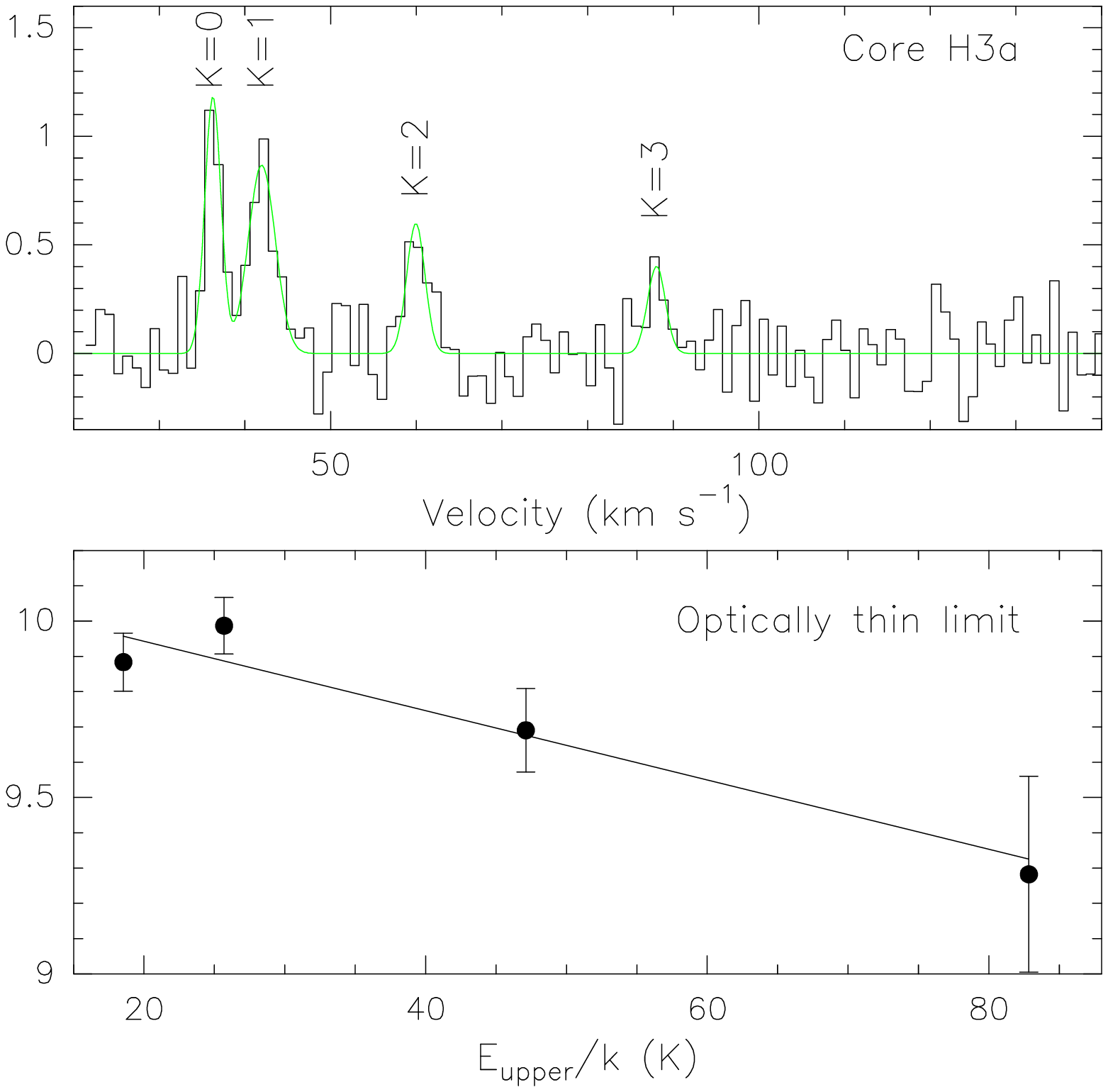}
\includegraphics[angle=-90,width=5.5cm]{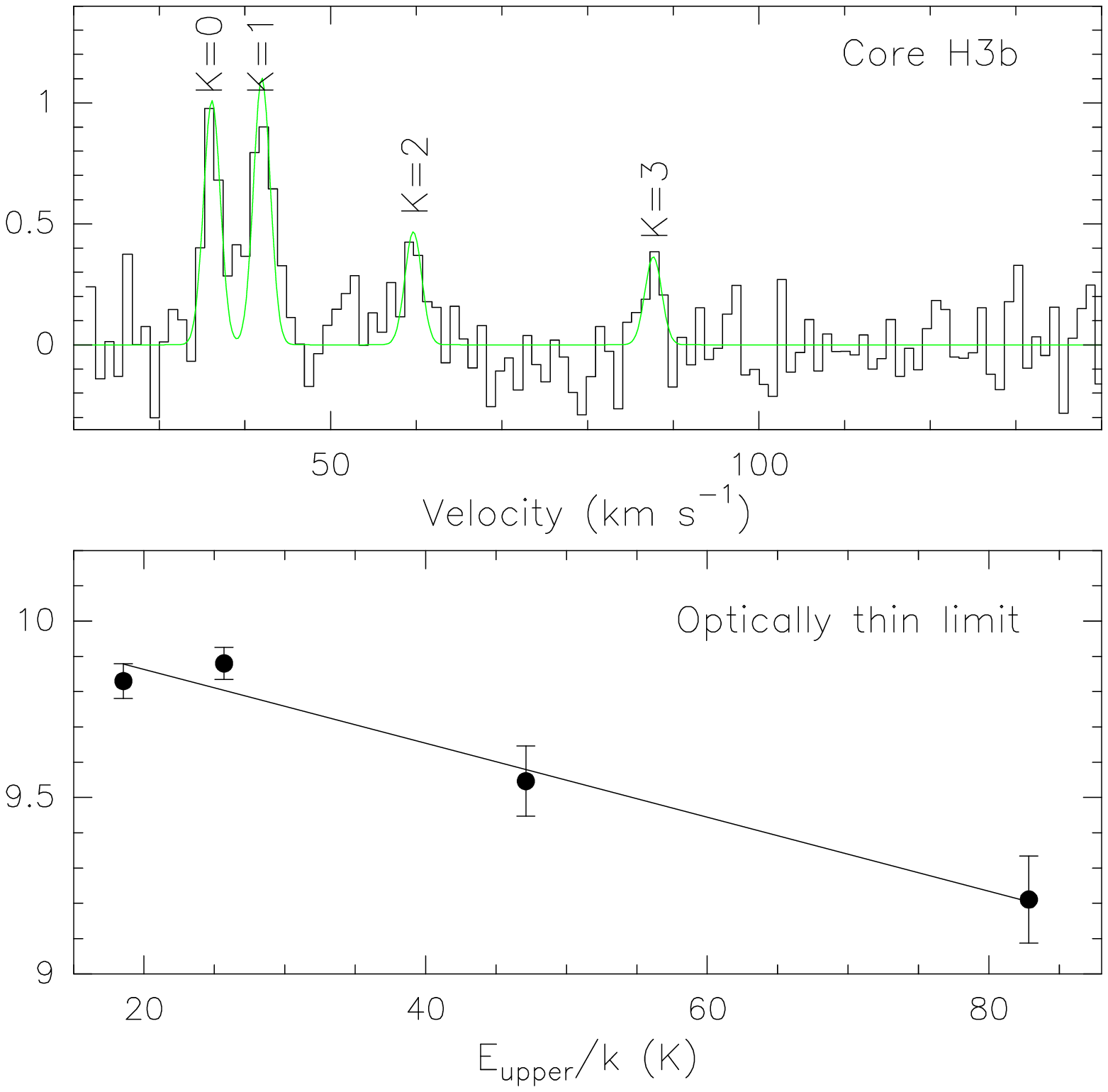}
\caption{{Upper panels} show the spectrum of the CH$_{3}$CN $K$=0,1,2,3 transitions and the fit (in green), while the {\em lower panels} show the \mecn\ rotational diagram for cores H-2a, H-3a, and H-3b. The rest frequency has been set on the $K$=1 level.}
\label{fig:t_rot3} 
\end{figure*}

\mecn\ is a commonly observed molecule in hot molecular cores, which are regions of high-density molecular gas surrounding an intermediate to high-mass YSO. The YSO elevates the dust temperatures ($>100$\,K) in these regions, which causes the evaporation of the icy grain mantles and the increase of the \mecn\ gas phase abundance by several orders of magnitude (\citealt{olmi:1996}; \citealt{tak:2000}; \citealt{purcell:2006}). The \mecn\ observations had a high angular resolution (see Table \ref{ta:spec}) allowing us to observe core-sized ($<$0.1pc radius) objects and to find the first signs of fragmentation in the clumps detected with {\em Herschel}. The integrated intensity \mecn\ map in Fig.~\ref{fig:maser_ch3cn} shows that CH$_{3}$CN is concentrated toward three cores, whose continuum properties are described in Section~\ref{sec:compact_sources}. The most compact core H-2a, which appears to be unresolved, overlaps with clump H-2, while clump H-3 was found to fragment into two \mecn\ cores, H-3a (no continuum emission detected) and H-3b (continuum emission detected). It is probable that the \mecn\ emission at the location of H-3a is extended emission from low velocity shocks, such as is seen in DR\,21 (\citealt{csengeri:2011b}) where it was associated with small-scale convergence flows, rather than the \mecn\ emission associated with a YSO. Such a nature of the \mecn\ emission is plausible, as core H-3a coincides with the red CO(2--1) outflow lobe, which could provide the low-velocity shocks (the outflow is discussed in Section~\ref{sec:outflow}). No \mecn\ emission was found toward clump H-1. 

Spectra integrated over the area of the cores were extracted from the data cube and fitted with GILDAS/CLASS\footnote{http://www.iram.fr/IRAMFR/GILDAS} software. The $K$=0 to $K$=3 rotational ladder components were detected for all three cores and fitted simultaneously with Gaussian profiles. We constrained the line width of all $K$ levels to be the same, i.e. that of the $K$=0 component which is given in Table~\ref{tab:ch3cn}. The signal-to-noise ratio for the $K$=4 component was too low to obtain a reliable fit. The temperature and column density were estimated using the rotational diagram analysis method (see, e.g., \citealt{araya:2005}) assuming that the regions are optically thin (\citealt{turner:1991}). In the upper panel of Fig.~\ref{fig:t_rot3} we show an example of the \mecn\ $K=0,1,2,3$ component spectrum and its fit. The bottom panel shows the population diagram for the various $K$ levels on which the temperature and column density estimation was based. The positions of the cores and results of the fits are given in Table~\ref{tab:ch3cn}. 

The optical depth can be calculated once the rotational temperature is known. If the \mecn\ emission is optically thick, the column densities will be underestimated, while the rotational temperatures will be over estimated (\citealt{goldsmith:1999}). We derived optical depths of the order of 0.02. Iterating by the optical depth correction factor ($\tau/(1-\exp^{-\tau})$), as done by \citet{remijan:2004}, we estimated that the corrected column densities and rotation temperatures are the same as those for the optically thin limit within the uncertainties. The calculated \mecn\ column densities of $\sim$$8\times10^{12}$\,cm$^{-2}$ (Table \ref{tab:ch3cn}) are similar to those found in hot cores (\citealt{olmi:1993,araya:2005}). The rotation temperatures show that core H-2a is warmer than cores H-3a and b, and in addition is also more compact (Table \ref{tab:ch3cn}). The \mecn\ line thermal widths are all dominated by non-thermal components (thermal \mecn\ line widths for 40--60\,K are $\sim$0.2--0.3\,\kms). 

\begin{table*}
\centering
\begin{minipage}{180cm}
\caption{Ammonia properties of clumps and cores}
\label{tab:nh3}
\begin{tabular}{lccccccccc}
\hline
Object  & $T_\mathrm{mb}^{11}$& $\Delta v^{11}$&$V_\mathrm{LSR}^{11}$& $\tau^{11}_m$&  $T_\mathrm{mb}^{22}$& $\Delta v^{22}$&$V_\mathrm{LSR}^{22}$  &T$_\mathrm{rot}$  & $N_\mathrm{NH_3}$\\ 
             &(K) & (km~s$^{-1}$) &(km~s$^{-1}$) &  &(K) &(\kms ) & (\kms ) &(K)&(10$^{15}$cm$^{-2}$) \\ \hline 
clump H-1 & --0.22$\pm$0.10 & 1.87$\pm$0.18 & 42.92$\pm$0.08 & 1.74$\pm$0.67 & --0.11$\pm$0.02&2.18$\pm$0.37&43.04$\pm$0.17&18$\pm$7 & 3.9$\pm$2.2\\
clump H-2$_1$ & 0.19$\pm$0.08 & 1.34$\pm$0.16 & 42.99$\pm$0.08 &1.30$\pm$0.5$^a$& 0.18$\pm$0.03 & 2.00$\pm$0.26 & 42.84$\pm$0.11 & 40$\pm$41 & -- \\
clump H-2$_2$& 0.22$\pm$0.09 & 1.26$\pm$0.13 & 45.60$\pm$0.06 &1.3$\pm$0.50$^a$ & 0.14$\pm$0.03 & 1.74$\pm$0.30 &45.47$\pm$0.13 & 24$\pm$13 & -- \\
clump H-3   & 0.65$\pm$0.07 & 1.84$\pm$0.06 &42.23$\pm$0.02 &2.47$\pm$0.24 &0.47$\pm$0.03&2.26$\pm$0.10 &42.17$\pm$0.04 & 22$\pm$3 &6.3$\pm$1.1\\
core H-2a$_1$ & 1.17$\pm$9.57 & 1.49$\pm$0.28 & 42.87$\pm$0.10 & 0.22$\pm$1.80&0.24$\pm$0.05&1.73$\pm$0.27&42.78$\pm$0.10&-- &-- \\
core H-2a$_2$ & -- & -- & -- & -- & 0.10$\pm$0.05 & 1.59$\pm$0.63& 45.26$\pm$0.29&-- &--\\
core H-3a  &  0.72$\pm$0.08 & 1.68$\pm$0.05 & 42.17$\pm$0.02 & 2.47$\pm$0.22 & 0.52$\pm$0.03 &2.13$\pm$0.10 &42.13$\pm$0.04 &22$\pm$3.2&6$\pm$1.0\\
core H-3b & 0.75$\pm$0.09 & 1.79$\pm$0.06 & 42.18$\pm$0.03 &2.81$\pm$0.28 & 0.58$\pm$0.04 & 2.25$\pm$0.11 & 42.15$\pm$0.04 & 23$\pm$3.9 &7$\pm$1.4\\
 \hline
\end{tabular} 

NOTES. Columns are (from left to right): name of object; \amm\,(1,1) main beam brightness temperature; \amm\,(1,1) line width;  \amm\,(1,1) LSR velocity; \amm\,(1,1) \\main group opacity; \amm\,(2,2) main beam brightness temperature; \amm\,(2,2) line width;  \amm\,(2,2) LSR velocity; \amm\ rotational temperature; \amm\ column density\\  

\end{minipage} 
\end{table*}

\begin{figure*}
\includegraphics[angle=-90,width=18cm]{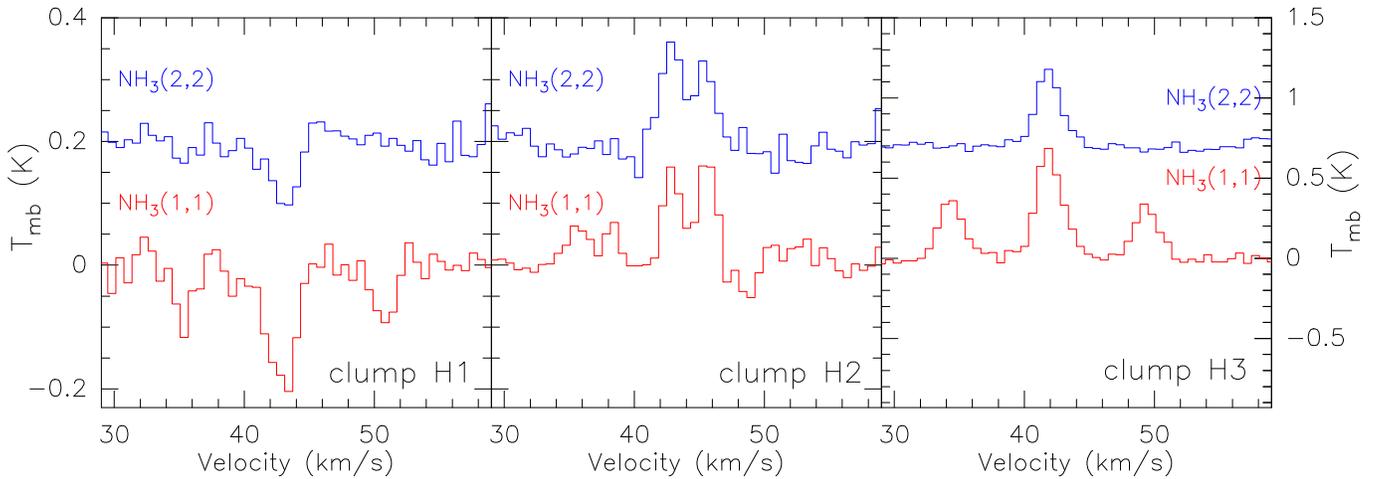}
\caption{\amm\,(1,1) and (2,2) spectra of clumps H-1, H-2, and H-3. The left-side y-axis gives the main beam temperature scale for the clump H-1 and H-2 spectra, while the right-side y-axis gives the main beam temperature scale for the clump H-3 spectra. The \amm\,(2,2) spectra have been offset artificially.}
\label{fig:amm_spec} 
\end{figure*}

\subsection{Ammonia (\amm) emission}
\label{sec:nh3}

The \amm\,(1,1) and (2,2) transitions are excited in dense and cold environments (e.g., \citealt{pillai:2006}). We analysed the \amm\ data cubes with the GILDAS software fitting the hyperfine structure of the \amm(1,1) emission and using a gaussian for the \amm(2,2) emission. The \amm\ spectra were extracted in the same areas as defined by the continuum clumps and the \mecn\ cores enabling a direct comparison between \amm, dust, and \mecn\ emission. Following \citet{rygl:2010b} we then derived \amm\ rotational temperatures and \amm\ column densities of the clumps and cores (see Table \ref{tab:nh3}). As the column densities were calculated based on interferometric observations, we might filter out some of the \amm\ emission, hence the \amm\ column densities might be higher than the values we found.

Figure \ref{fig:amm_spec} shows the \amm\ spectra of the clumps. Towards clump H-1 we found \amm\ absorption, similar to that toward DR\,21 (\citealt{cyganowski:2003}), indicating that the emission is emanating from cool molecular material in front of the 10$^3$\,K hot  \hiir. The \amm\ rotational temperature of clump H-1, 17.6\,K was much lower than the dust temperature (42\,K), which is a possible indication of the \amm\ being more shielded from the \hiir\ radiation than the dust. \amm\ emission toward clump H-2 was found to have two components (around 43\,\kms and 45.6\,\kms), making the fitting of the lines more difficult. We fixed the optical depth (and its error) at 1.2$\pm$0.5 for both components when performing the line fit. The two velocity components are also found in the \c18o\ spectra (see Section \ref{sec:c18o}), where the 43\,\kms\ component is correlated with the ridge emission, and the 45\,\kms\ component is thought to emanate from the arc-like structure around the \hiir . Unfortunately, the temperature estimations have too large uncertainties to provide more detailed information on these two velocity components. 

Clump H-3 had the strongest \amm\ emission among the three clumps. Its \amm\ rotational temperature (22\,K) was found to be in very good agreement with the dust temperature (20\,K). The \amm\ properties of clump H-3, which overlaps with \citet{pillai:2011} \amm\ core no.~3, are in agreement with the rotational temperatures and \amm\ column densities derived by those authors. 
The \mecn\ rotational temperatures were higher than the \amm\ temperatures, which is expected since \mecn\ traces warm regions near the YSO, while \amm(1,1) and (2,2) trace colder gas. 
The \amm\ analysis of core H-2a yielded similar numbers as for clump H-2, except that for the \amm\,(1,1) line we did not detected the 45.6\.\kms\ component due a higher noise in the core spectrum. 
Similarly, core H-3a and b have similar \amm\ properties as clump H-3. The LSR velocities of the \amm\ toward the cores H-2a, H-3a, and H-3b match well with the \mecn\ velocities, indicating that both molecules trace material with a similar kinematic signature, belonging to the same object.  

While the \amm(1,1) line width is an intrinsic line width (since we fitted the hyperfine structure of the transition), the \amm(2,2) width is not: it is an overestimate of the intrinsic line width. For both \amm\ and \mecn , the line widths are dominated by the non-thermal line width, since thermal line widths for \amm\ are of the order of 0.2--0.3\,\kms\ for temperatures of 20--30\,K. For all cores, the \amm\,(1,1) and (2,2) line widths were both much broader than those of \mecn\ ($\sim$1--1.2\,\kms). Possibly, the \mecn\ emission emanates from a smaller area near the YSO, while the \amm\ emission comes from a larger volume containing more turbulence in the surrounding medium by possible disks and outflows (see the outflow in clump H-3 in Section~\ref{sec:outflow}).

\subsection{Outflows}
\label{sec:outflow}

\begin{figure}
\includegraphics[width=10cm,angle=-90]{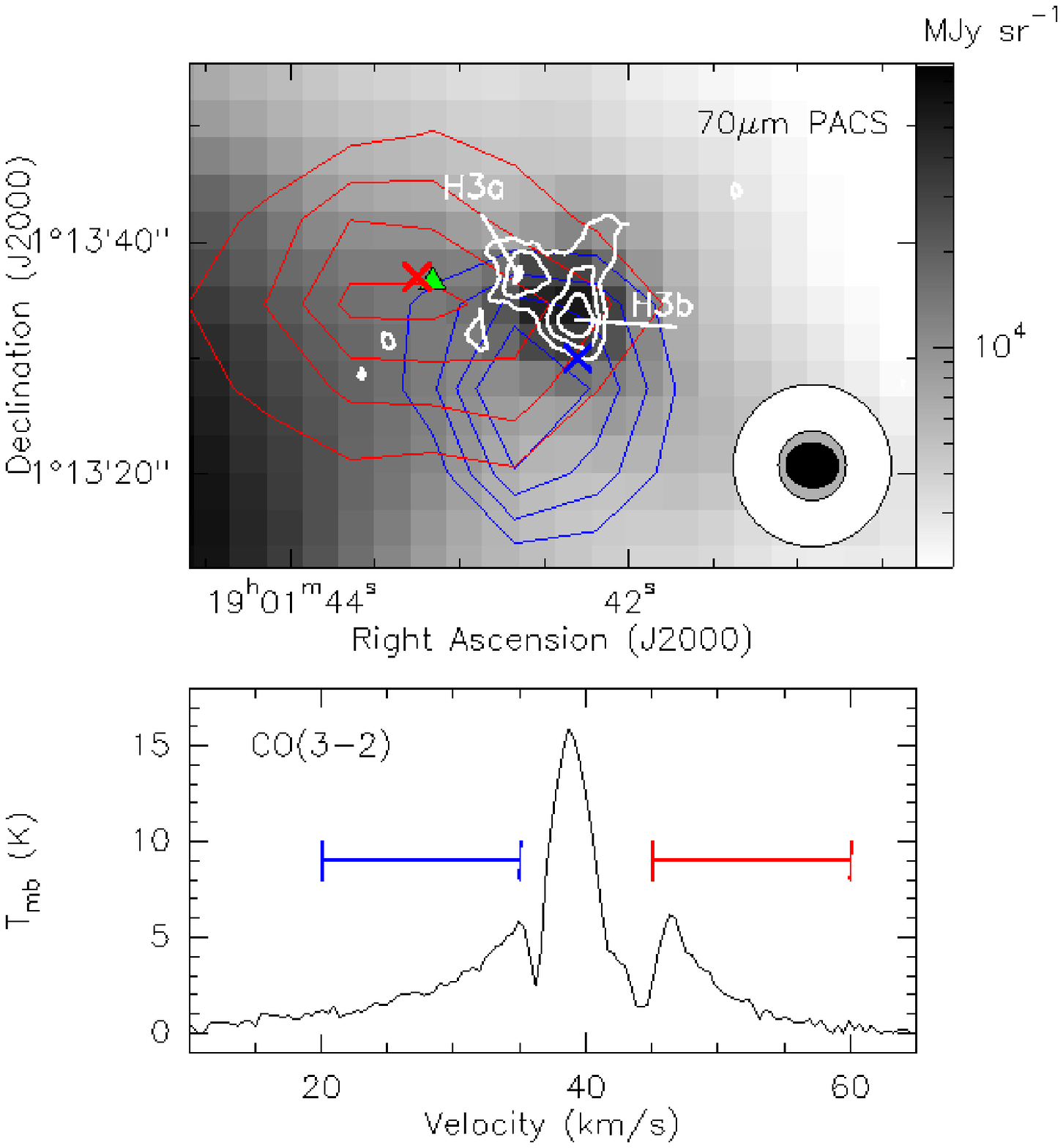}
\caption{\label{fig:outflow} {\em Top panel:} 70\um\ PACS greyscale map centred on clump H-3 with the JCMT CO\,(3--2) outflow (blue, red) and the \mecn\ cores (white) overlaid in contours. CO contour levels are 20, 30, 40, and 50 K\,\kms for both blue and red shifted material. BIMA \mecn\  contour levels are 0.06 0.08 0.10 0.12 Jy beam$^{-1}$\,\kms. Beam sizes of the CO\,(3--2) (white), PACS 70\,$\mu$m (grey), and \mecn\ emission (black) are given in bottom right corner. The green triangle represents the water maser location (\citealt{hofner:1996}), the blue and red crosses mark the HCO$^+$ outflow lobes from \citet{pillai:2011}. {\em Bottom panel:} the CO(3--2) spectrum toward clump H-3 where the velocity ranges used for mapping of the blue and red shifted outflow lobes are indicated.}
\end{figure}

Molecular outflows are commonly observed in low-mass and high-mass star-forming regions (see, e.g., \citealt{phillips:1988,beuther:2002,sepulcre:2009,rygl:2013b,duarte-cabral:2013}). Outflows arise from the mass accretion onto the molecular clump (e.g., \citealt{shepherd:1996}), and are often accompanied by maser emission due to the shocks created by the outflow (e.g., see the water masers in the outflow of W3(OH) in \citealt{hachisuka:2006}). Molecular outflows can be observed through optically thick emission of common ISM constituents, such as $^{12}$CO.

The HARP CO\,(3--2) data revealed an outflow in clump H-3. In Fig.~\ref{fig:outflow} we show the integrated emission of the high-velocity wings. The blue and red high-velocity lobes are quite compact indicating that the outflowing material is collimated, and that clump H-3 must be hosting active star-formation activity. A comparison with 70\,$\mu$m emission, tracing mainly warm dust near YSOs, shows that core H-3b (rather \mecn\ core H-3a) is the more likely origin for the outflow since it coincides with the 70\,$\mu$m source. The CO\,(3--2) outflow lobes coincide with the lower resolution \hcop(1--0) outflow lobes discussed by \citet{pillai:2011} (see Fig.~\ref{fig:outflow}). The displacement between the CO outflow lobes and the {\em Herschel} dust continuum is real, since it is greater than the sum of the position uncertainties of the JCMT CO data (2\arcsec) and BIMA \mecn\ data (0$\rlap{.}$\arcsec4). Also the \hcop(1--0) outflow and the 3\,mm continuum emission in \citet{pillai:2011} show the same displacement. Apparently the outflow axis is not straight, which is why the blue and red lobe are not symmetrical around the continuum source. \citet{pillai:2011} estimated the outflow mass to be about 5\,\msun, indicating that core H-3b is driving a massive outflow. In addition to this massive molecular outflow, clump H-3 also hosts water masers (\citealt{hofner:1996}) located toward the red outflow lobe. The detection of the massive outflow and water masers indicates that clump H-3 is in an active star-formation stage, hence presumably younger than the other two clumps H-1 and H-2, which do not exhibit any collimated outflows.

\section{Results and analysis of extended emission}
\subsection{{\em Herschel} continuum emission}
\label{sec:extended_herschel}

We computed column density and temperature maps from the calibrated 160--500\,$\mu$m images, after these were convolved to the 500\,$\mu$m resolution (36\arcsec) and rebinned to the pixel size of the 500\,$\mu$m map (11$\rlap{.}$\arcsec 5). Then, pixel-by-pixel modified black body fits were performed on the regridded maps (see \citealt{elia:2013} for more the expression and its details).
We assumed a dust opacity $\kappa_\nu$ of $\kappa_\nu=\kappa_0\big(\frac{\nu}{\nu_0}\big)^\beta$ with $\kappa_0$=$0.1\,\mathrm{cm^2 g^{-1}}$ (which includes a gas to dust ratio of 100) and $\nu_0$=1200\,GHz (250\,$\mu$m) (\citealt{hildebrand:1983}). The dust emissivity index $\beta$ was kept constant at 2.0. Recent studies (e.g., \citealt{juvela:2011,miettinen:2012,sadavoy:2013}) show that $\beta$ varies across the cloud. While some works point out the anti-correlation between dust temperature and $\beta$, Bayesian SED-fitting finds a anti-correlation of $\beta$ with column density rather than temperature (\citealt{kelly:2012}). \citet{ysard:2013} find that the change in $\beta$ is due to dust-properties which vary with density. With our {\em Herschel}-only data (the SCUBA, MAMBO and BIMA dust continuum maps had a much smaller coverage, so were not used for the large-scale temperature and column density maps) are not able to constrain the $\beta$. Using $\beta=2.0$ with $\chi^2$ fitting (as we do) can cause the resulting temperature to be too low in case $\beta<2.0$ and vice versa, as shown by \citet{sadavoy:2013}. Furthermore, to obtain the H$_2$ column density we used the mean molecular weight per hydrogen atom \,$\mu_\mathrm{H}$=2.8  (\citealt{kauffmann:2008}). Recent HOBYS studies, such as e.g. \citet{hill:2012}, often use a mean molecular weight of 2.3 which is averaged per particle instead per hydrogen atom and results in higher column densities by a fixed factor of 2.8/2.3$\sim$1.2.

\begin{figure}
\centering
\includegraphics[angle=-90,width=8cm]{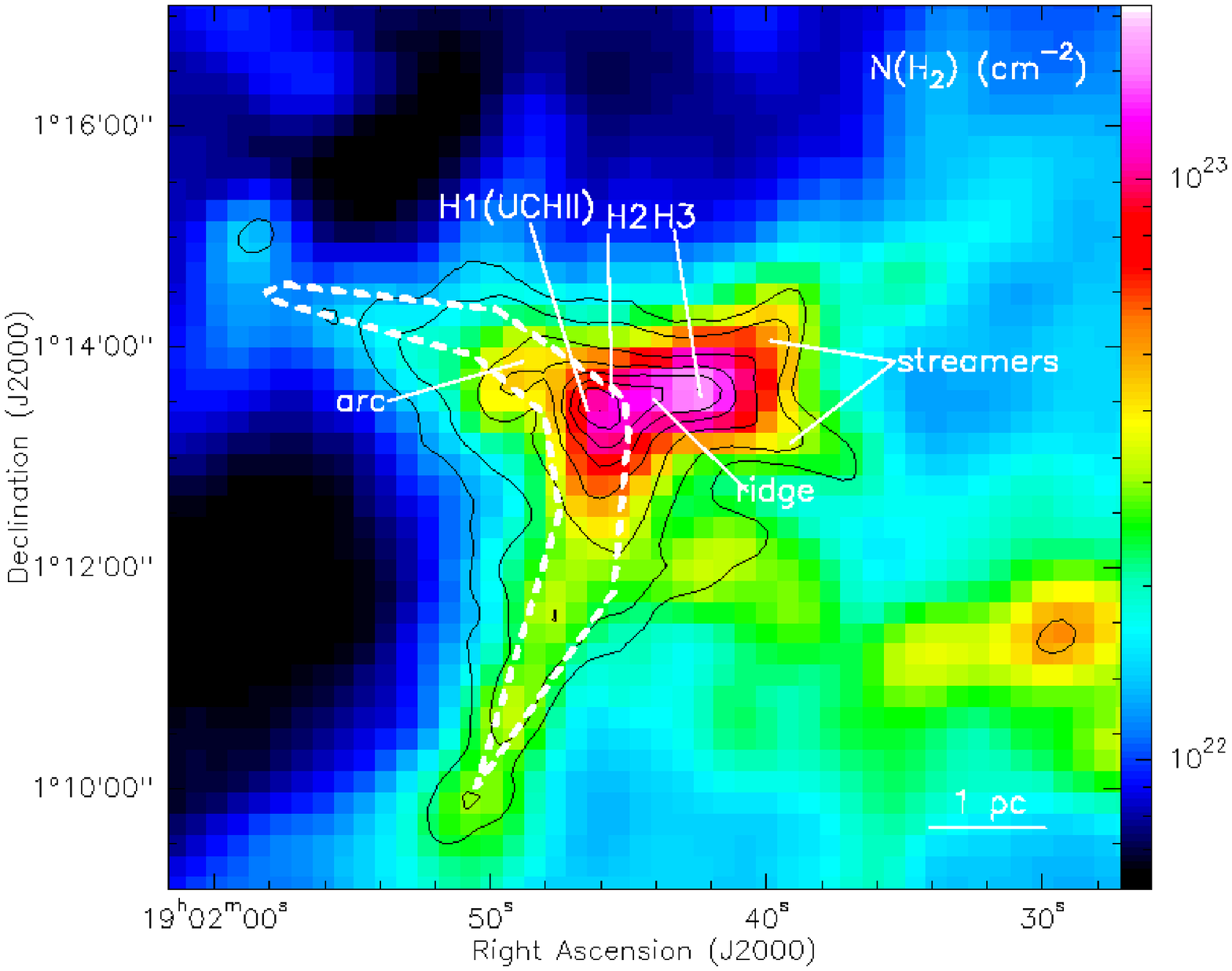}
\includegraphics[angle=-90,width=8cm]{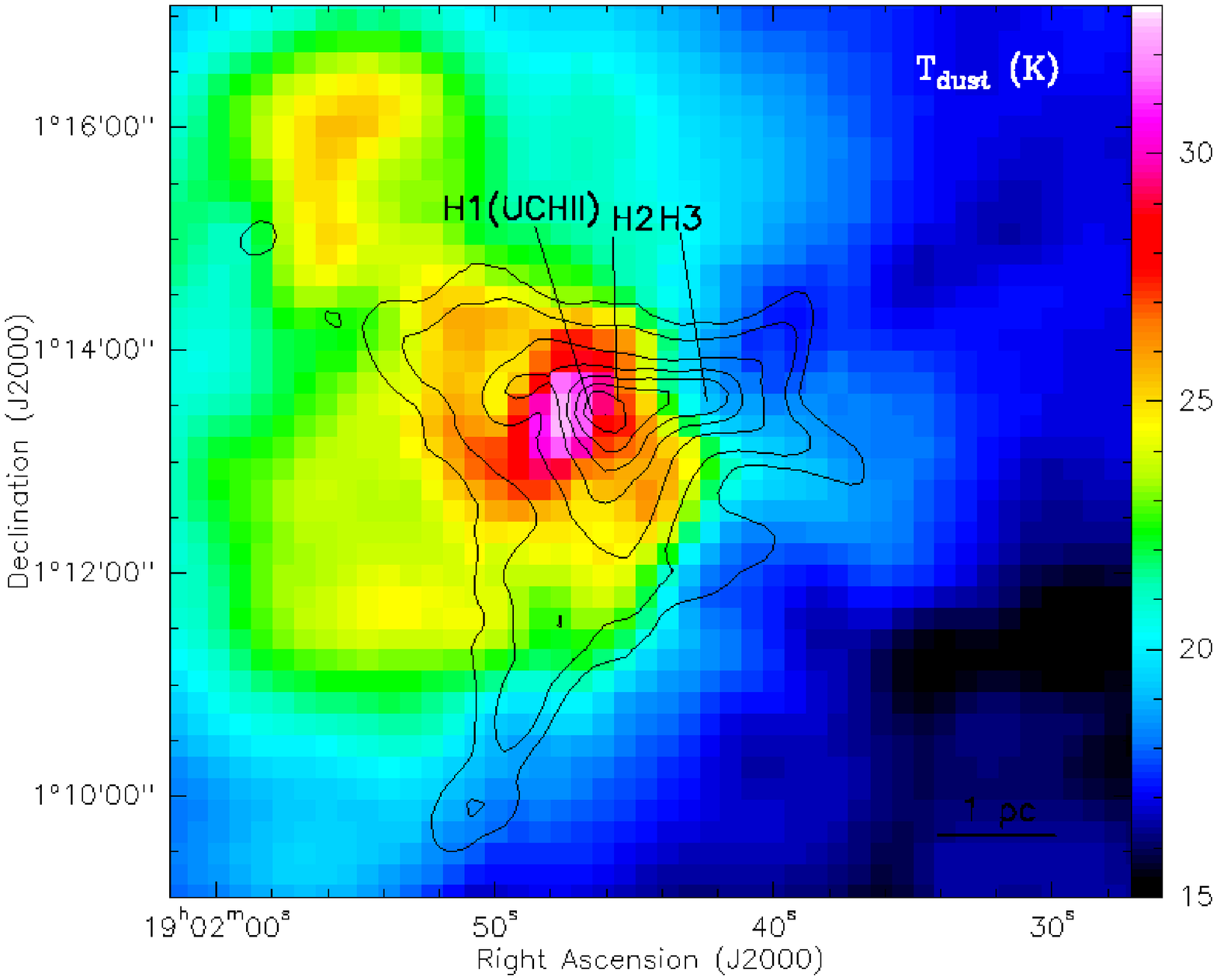}
\caption{\label{fig:ntmaps_small} {\em Herschel} H$_2$ column density and dust temperature maps of the W\,48A surroundings. Clumps H-1, H-2, H-3 are marked, just as the large scale structures: the arc (white dash), the ridge and the streamers. Contours show the SPIRE 250\,$\mu$m emission between 2000 and 55000 MJy\,sr$^{-1}$ using a square root scaling.}
\end{figure}

Figure~\ref{fig:ntmaps_small} displays the resulting temperature and column density maps of W\,48A with the 250\,$\mu$m contours overplotted. As expected, the 250\,$\mu$m contours match very well with the column density map, but are quite different from the temperature map, which traces the dust heated by YSOs and stars. The column density map shows: a north-west oriented arc around the \hiir\ (dashed contours, Fig.~\ref{fig:ntmaps_small}) directed toward a dense ridge\footnote{The word ridge is used to indicate an filamentary cloud with a column density $>10^{23}\,\mathrm{cm^{-2}}$ (see \citealt{hill:2011}).} running east-west from clump H-1 till clump H-3, and diverging into two ``cold streamers", westwards of clump H-3. The morphology of the arc suggests that it was shaped by the \hiir . 
The temperature map (Fig.~\ref{fig:ntmaps_small}) shows how the W\,48A \hiir\ is heating up the surroundings. Heating of dust by (young) stars has been seen in many {\em Herschel} observations (see, e.g., \citealt{zavagno:2010,hill:2012,minier:2013}).
Towards the low-density region, south-east of the \hiir\ there is a bubble of heated dust ($\geq$25\,K) with respect to the background temperature ($\sim$16\,K), while toward the more dense material the heating is less efficient: the ridge (T$\leq$26\,K), and especially the streamers (T$\leq$20\,K) are composed of colder dust than the \hiir\ ($\sim$28\,K). 
With a molecule like \amm\ we can use the ratio of the (2,2) to (1,1) main beam temperatures ($T_{22}/T_{11}$) to find heated molecular gas (\citealt{torrelles:1989,zhang:2002}). The VLA \amm\ data find the highest $T_{22}/T_{11}$ near core H-2 (see Fig.~\ref{fig:ammratio}), suggesting that it might be heated by the \hiir , and consistent with the presence of a PDR as seen near other \hiir\ regions (e.g., \citealt{palau:2007}).

The cold streamers seen by {\em Herschel} were first noted by \citet{pillai:2011}, who found several NH$_2$D cores just westward of clump H-3 using interferometric PdBI data. To deuterate \amm, creating the NH$_2$D molecule, very cold temperatures (T$<$15\,K, \citealt{bergin:2007}) are required. Protostellar heating of the environment, which would increase the temperature T$>$20\,K would destroy the NH$_2$D. Therefore the detection of NH$_2$D cores indicates that in this region no protostars have been formed yet. These NH$_2$D cores were not detected in the {\em Herschel} maps, nor in the high resolution 1.3\,mm PdBI data of \citet{pillai:2011}, showing that at small core-like scales the deuterated gas differs from the dust distribution, while on larger scales (as 1\,pc, such as the streamers) the correlation between the dust and NH$_2$D emission is remarkably good. The {\em Herschel} dust temperature and column density map shows that these streamers have indeed the lowest dust temperature of the whole W\,48A environment and that they are dense structures. We conclude that the cold streamers represent the earliest star-formation stage observed in the W\,48A star-forming region. 

\subsection{Kinematics}
\subsubsection{\c18o\,(1--0) gas kinematics near the UC H{\sc ii} region}
\label{sec:c18o}

\begin{figure*}
\includegraphics[angle=-90,width=8cm]{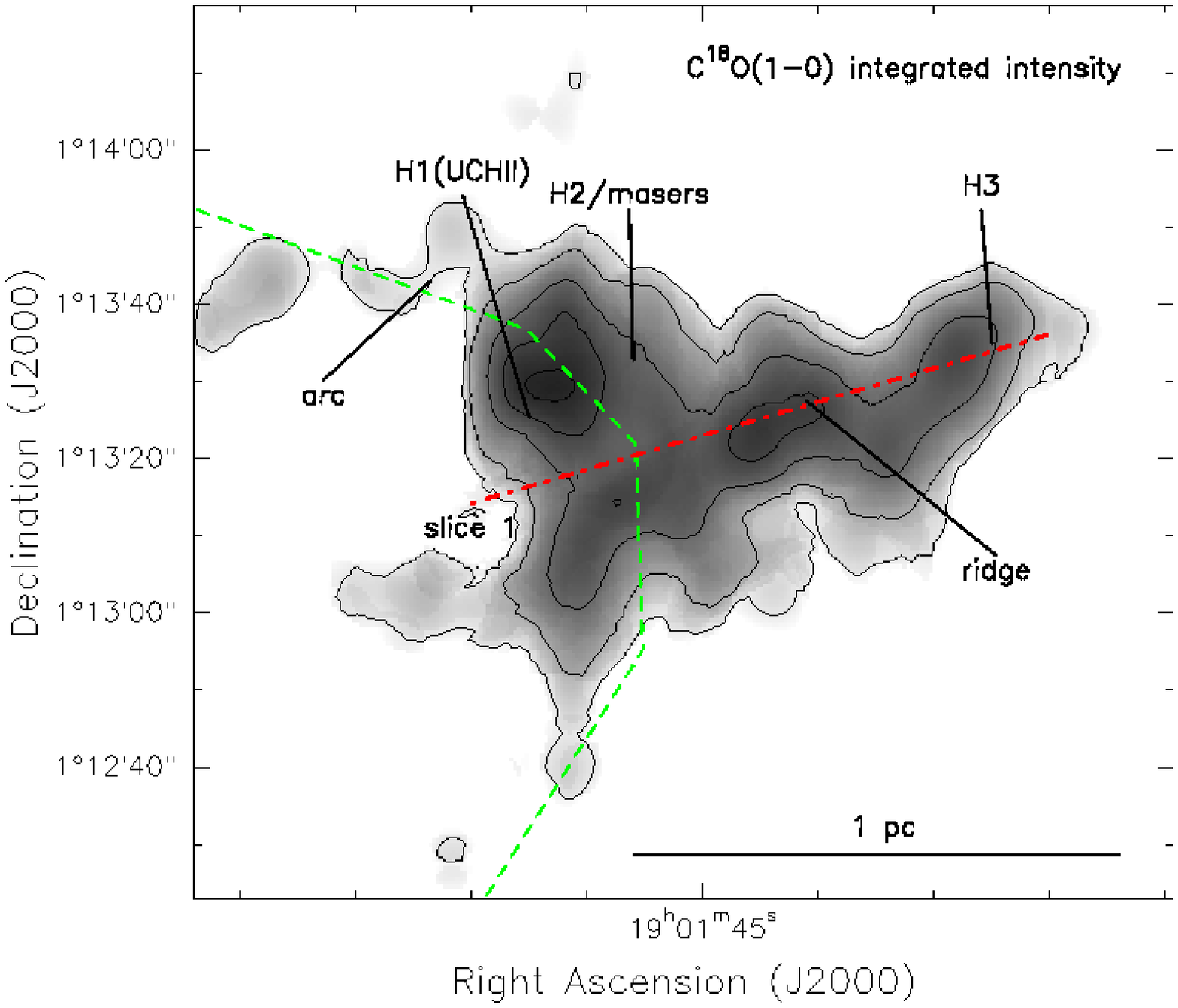}
\includegraphics[angle=-90,width=6.9cm]{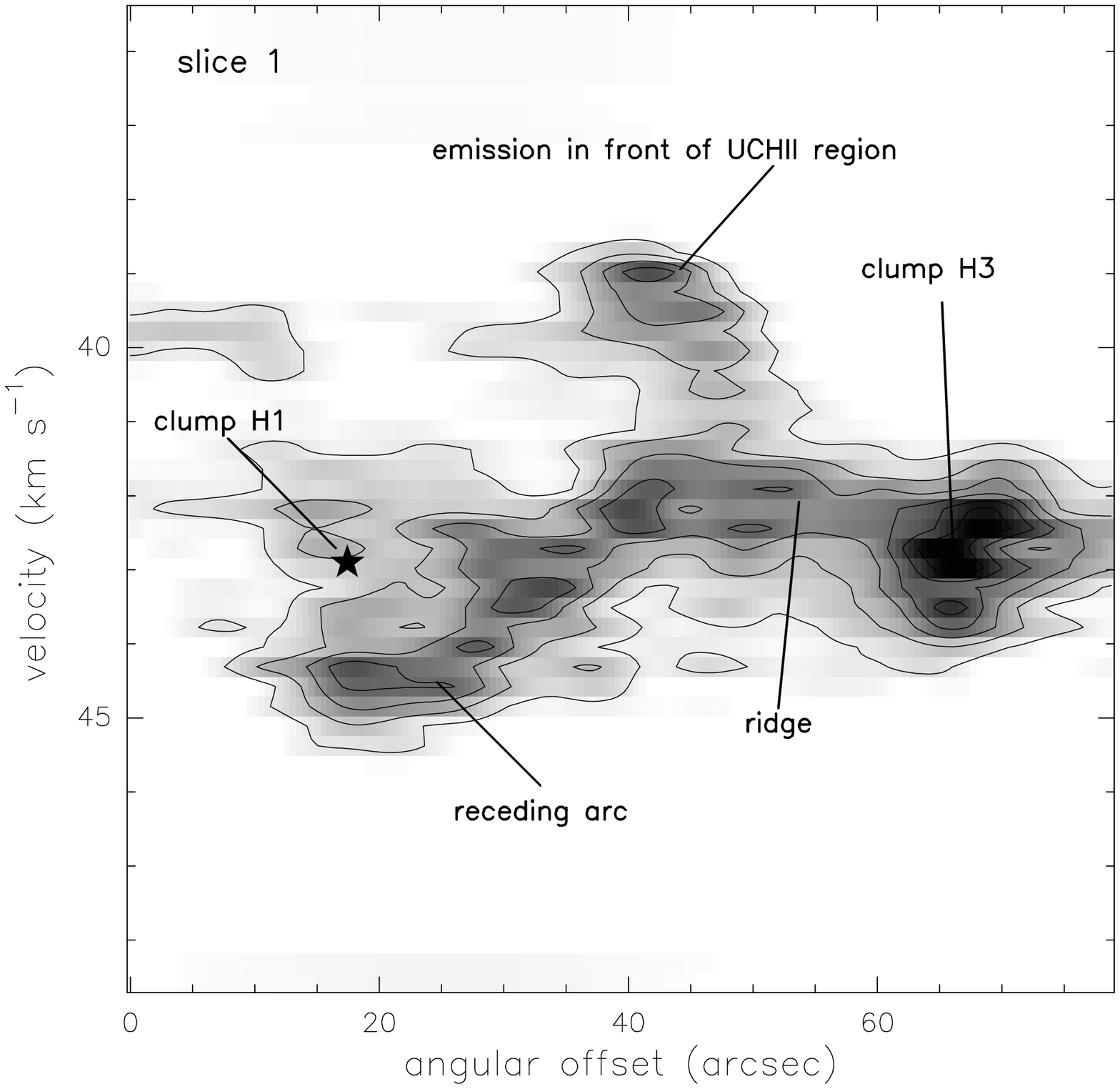}
\caption{{\em Left panel:} BIMA+IRAM\,30m integrated intensity map of the \c18o\ emission. Clump H-1 (W\,48A \hiir), core H-2a (maser source and clump H-2), clump H-3, the ridge, and the arc are marked. Contours are 0.25, 0.5, 1.0, 1.5, 2.0\,Jy\,beam$^{-1}$\,\kms. {\em Right panel:} Position velocity plot along the red dash-dot line in the left figure. Contours are 0.15 0.25 0.35 0.45 0.55 0.65 0.75 mJy beam$^{-1}$. The various velocity components are marked on the plot. \label{fig:co_mom0}}
\end{figure*}

\begin{figure*}
\includegraphics[angle=-90,width=12cm]{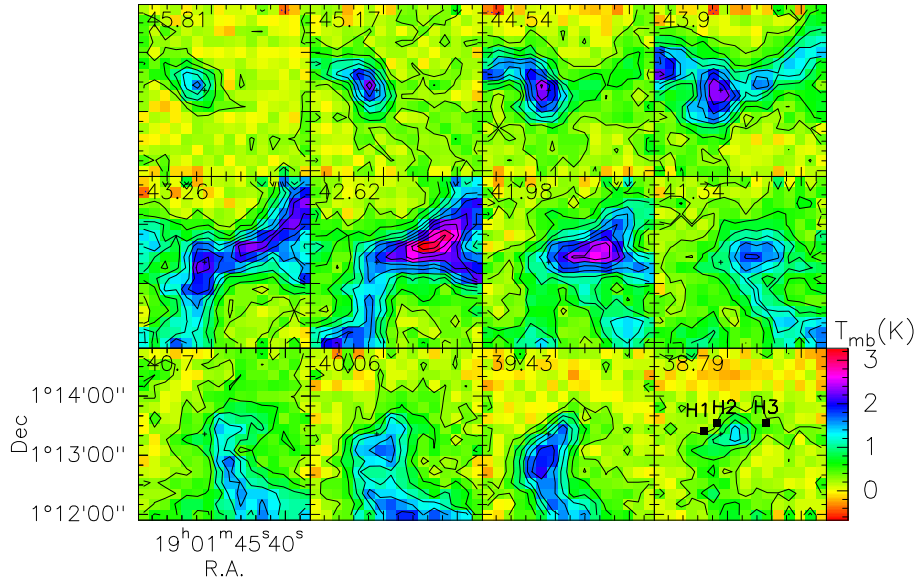}
\caption{IRAM\,30m \c18o\,(1--0) channel map smoothed to a channel spacing of 0.64\,\kms\ for a better visualisation. Central velocity of each channel is marked in the top left corner. The map noise is 0.13\,K and the contour spacing is 0.3\,K (2.3\,$\sigma$). Black squares indicate clumps H-1, H-2 and H-3. \label{fig:chanmapc18o}}
\end{figure*}

We proceed with the exploration of the kinematics of the extended structures of the W\,48A star-forming region through the \c18o\,(1--0) data. The BIMA+IRAM30m \c18o\,(1--0) integrated intensity map 
(Fig.~\ref{fig:co_mom0}, left panel) shows a large area of emission around the head of the arc and a horizontal extension (the ridge) with several bends, that ends near the H-3 clump. The peak of the \c18o\,(1--0) emission is near the H{\sc ii} region. To investigate the gas kinematics, we made a \c18o\ position-velocity plot (right panel of Fig.~\ref{fig:co_mom0}) to show the \c18o\,(1--0) intensities at each velocity bin along the cut across the arc  and the ridge (indicated in the left panel). The ridge has a quite simple velocity structure: a very narrow velocity range (42$<V_\mathrm{LSR}<$43\,\kms) including the \mecn\ LSR velocities along the ridge and is considered to be ``at rest". Near the \hiir\ the kinematics become more interesting. There is receding arc at angular offsets $<40$\arcsec, velocities 42--45\,\kms\ and some material in front of the \hiir\ (at 40\arcsec). Together they might create a shell structure that is similar to the expanding bubbles in the model of \citet{arce:2011}. However, unlike in \citet{arce:2011} our source is not a bubble, but an arc or a shell. Also there is more material redshifted than blueshifted, indicating that we are not dealing with a symmetrical structure, and making the word ``arc" more appropriate than ``shell". Fig.~\ref{fig:co_mom0} suggests that the \hiir\ formed on the near side of the ridge, and during its evolution pushed most of the dense gas away to be redshifted. A smaller amount of material surrounding the \hiir\ ended up blueshifted or at rest.

\begin{figure*}
\centering
\includegraphics[angle=-90,width=12cm]{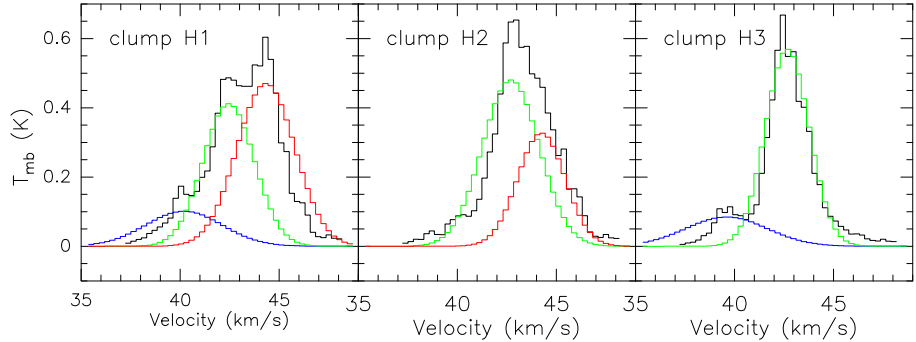}
\caption{\label{fig:c18o_spec} BIMA+IRAM 30m \c18o\,(1--0) spectra toward clumps H-1, H-2, and H-3, showing the Gaussian fits of the three velocity components in different colours.}
\end{figure*}

To visualise the multiple velocity components we use the IRAM 30m \c18o\,(1--0) channel map (Fig.~\ref{fig:chanmapc18o}) as the single dish \c18o\ data covered a larger area (the \c18o\ BIMA data were limited by the field of view). The \c18o\ emission seems to be a mixture of three main velocity components, one around 38--41\,\kms\ (which we name the ``blue" component, see bottom four panels in Fig.~\ref{fig:chanmapc18o}), one around 41--43.5\,\kms\ (the ``green" component, the middle four panels in Fig.~\ref{fig:chanmapc18o}), and one around 44--45\,\kms\ (the ``red'' component, top four panels in Fig.~\ref{fig:chanmapc18o}). 
The arc is clearly visible: starting south-east of clump H-1 (\hiir) at $v_\mathrm{LSR}$=42.6\,\kms and proceeding northwards toward clump H-1 and from there north-eastwards at $v_\mathrm{LSR}$=44.5\,\kms, covering a velocity gradient of $\sim$2\,\kms . The green velocity component contains most of the \c18o\ emission: it correlates well with the dust continuum and the column density map: the ridge, the streamers, and the middle-southern part of the arc are all visible. This not surprising since this velocity range also includes the LSR velocities of the clumps and cores. What looked one arc in the dust continuum, seems to kinetically be more complicated: the middle-northern part of the arc is receding, and the southern part of the arc is at rest or advancing. This points either to two uncorrelated velocity structures or to an arc which is inclined to result in the observed perspective. 
We obtained the clump \c18o\ spectra from the high-dynamic range BIMA+IRAM 30m data and fitted a Gaussian to each velocity component (Fig.~\ref{fig:c18o_spec}). The green velocity component is found toward all three clumps, while the red component is found only toward H-1 and H-2. The blue component, seen toward clump H-1 and H-3, is always much weaker than the green or the red component.

\subsubsection{\thco(1--0) emission tracing material in front of the \hiir}

\begin{figure}
\includegraphics[angle=-90,width=8cm]{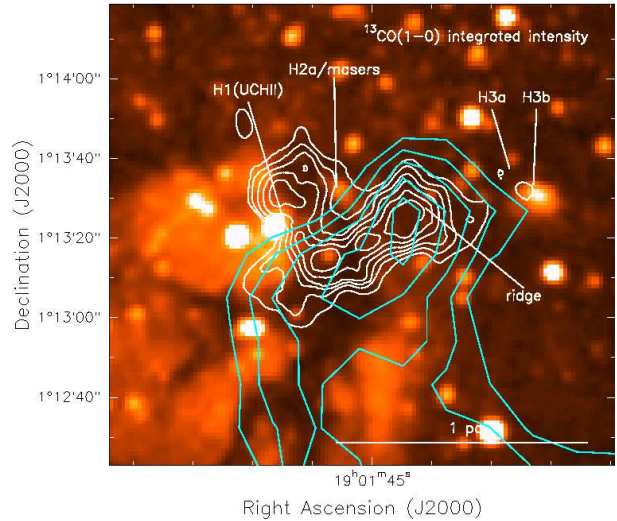}
\caption{2MASS Ks-band (2.16\,$\mu$m) image overlaid with \thco\ integrated intensity contours (white) and the blue \c18o\ velocity component (38--41\,\kms, cyan) . \label{fig:13co_kband}}
\end{figure}

This blue \c18o\ velocity component, introduced in the previous section, is different from the red and green \c18o\ components. Namely, while the red and green \c18o\ have a similar distribution to the dust emission, the blue \c18o\ does not. When one compares the $v_\mathrm{LSR}=38.79\,$\kms\ channel map in Fig.~\ref{fig:chanmapc18o} with the column density map in Fig.~\ref{fig:ntmaps_small}, one can notice that  none of the high-density structures are coinciding with the blue \c18o\ emission. In fact, the blue velocity component correlates well with the Ks-band absorption features, which are caused by material in front of the \hiir\ (Fig.~\ref{fig:13co_kband}). 

Our BIMA dataset included \thco\ observations. As \thco\ is about ten times more abundant in cold molecular clouds than \c18o\ (see e.g., \citealt{du:2008} and references therein), it is often observed to be optically thick, even where \c18o\ is still optically thin. Therefore, it can be used to visualise the location of low density molecular gas. If there was optically thick molecular gas in front of the \hiir , the \thco\ emission would have the same shape as the near-infrared absorption, as we indeed see in Fig.~\ref{fig:13co_kband}. The correlation of the blue \c18o\ component, the Ks absorption and the \thco\ emission suggests that it traces low density material in front of the \hiir . 

\section{Discussion}
\subsection{W\,48A revisited}

\citet{roshi:2005} sketched the W\,48A \hiir\ as a PDR dominated H{\sc ii} region, where the ionising star is moving with respect to the molecular cloud. Their C76$\alpha$ radio recombination line (RRL) measurements showed that the PDR ($V_\mathrm{LSR}$=41.9\,kms) was moving at 2.5$\pm$1\,\kms\ into the molecular cloud for which they took $V_\mathrm{LSR}$=44\,\kms\ (\citealt{churchwell:1992}). The LSR velocities of our \c18o\,(1--0) data show that the molecular cloud covers a range of velocities from $\sim$41 up to 45\,\kms . The LSR velocity of the C76$\alpha$ line falls in the velocity range of the molecular material: the PDR is moving {\em at the same velocity} as that part of the cloud. 

\begin{figure}
\includegraphics[angle=0,width=8cm]{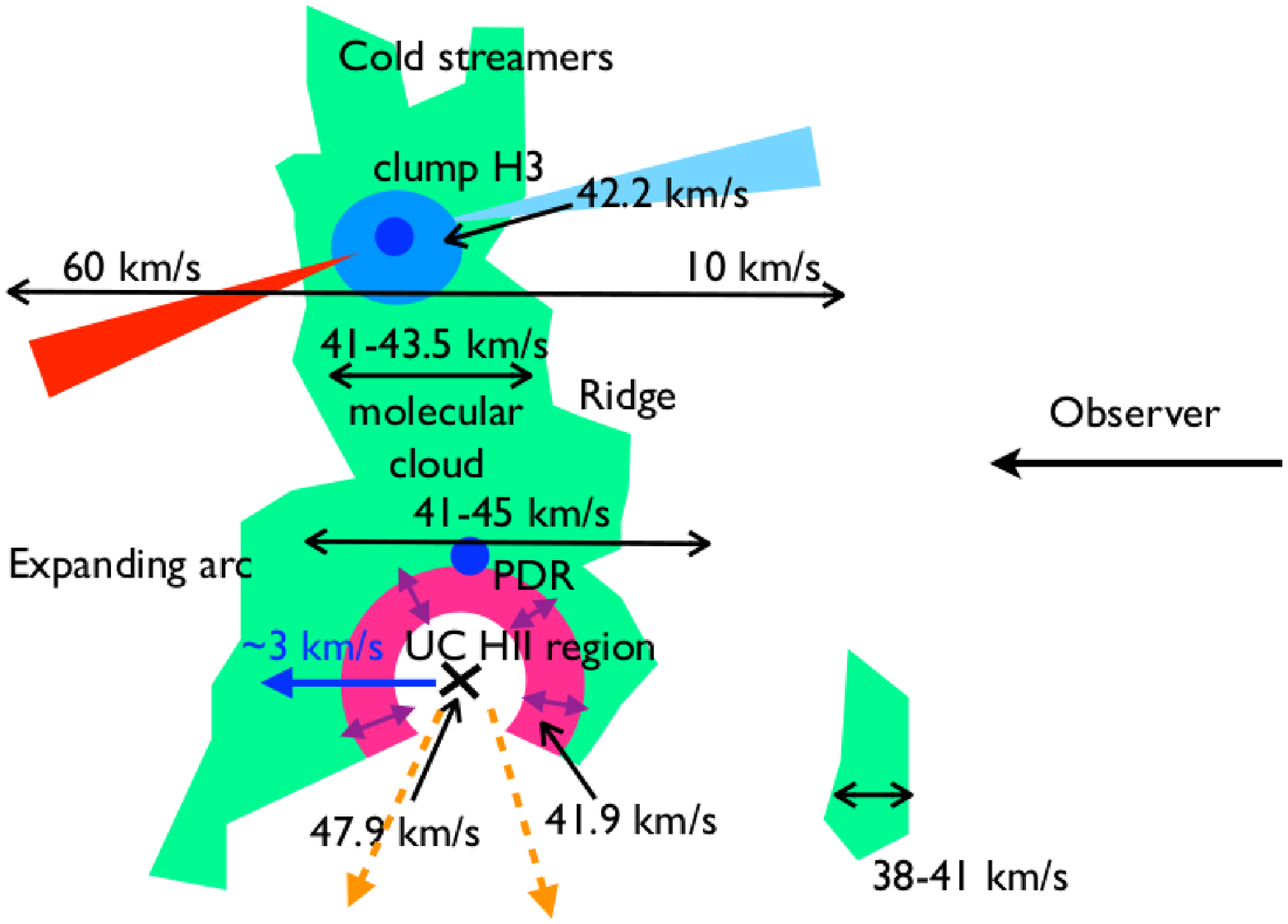}
\caption{\label{fig:sketch} Schematic drawing of the W\,48A star-forming region. The cross marks the location of the \hiir\ (white) with LSR velocity of 47.9\,\kms\ (\citealt{wood:1989a}). The PDR is pink and has an LSR velocity of 41.9\,\kms\ (\citealt{roshi:2005}). The molecular cloud is green and contains a wider range of velocities. The purple arrows in the PDR show that it is expanding. Core H2a is marked in dark blue at the border of the PDR. Light blue marks clump H-3 with the H-3b core inside it in dark blue. The molecular outflow of clump H-3 is shown by a red and blue cone indicating also the velocity range. The \hiir\ is moving away with about 3\,\kms . The champagne flow of the \hiir\ is indicated by dashed yellow arrows. }
\end{figure}

To understand if the PDR is pressure confined by the molecular material, we use the formulae given in \citet{roshi:2005} and calculate pressure of the molecular cloud. The molecular cloud pressure is given by the sum of the pressure from thermal processes and from turbulence (\citealt{xie:1996}). \citet{roshi:2005} demonstrated that this $P_\mathrm{PDR}=5.3-43.0\times10^{-7}\,\mathrm{dyne\,cm^{-2}}> P_\mathrm{UCHii}=1.3\times10^{-7}$\,dyne\,cm$^{-2}$ and that therefore the \hiir\ is pressure confined by the PDR. Now, the pressure in the molecular material in the ridge is given by
\begin{equation}
P_\mathrm{mol}= n_\mathrm{H_2} (k_BT_\mathrm{mol} +  \mu m_\mathrm{H} dv_\mathrm{mol}^2),
\end {equation} 
where $T_\mathrm{mol}=26\,$K (from the dust temperature map), $\mu=2.8$ and $dv_\mathrm{mol}=$1.0--1.8\,\kms\ based on \c18o\,(1--0) and \amm\,(1,1) line widths resulting in $P_\mathrm{mol}=n_\mathrm{H_2} 1\times10^{-13}$\,dyne\,cm$^{-2}$. Using the \amm\ rotational temperature (22\,K) would yield a similar outcome. Estimates of the volume density based on the column density map depend on the depth of the ridge. Assuming the same depth as width ($\sim$1\,pc), the volume density of the molecular material in the ridge would be $6\times10^4$\,cm$^{-3}$. This would bring the pressure of the molecular material to $6\times10^{-9}$\,dyne\,cm$^{-2}$, much lower than the PDR pressure. To stop the PDR from expanding into the molecular cloud, $P_\mathrm{mol}$ would have to be larger than $P_\mathrm{PDR}$ which would require molecular hydrogen densities of $10^8$\,cm$^{-3}$. However, our hydrogen density estimate for the ridge is several orders of magnitude smaller. Therefore, the PDR is expanding into the cloud and pushing the molecular material outward, maintaining the arc-like structure. To fully appreciate the role of the PDR, high-sensitivity RRL imaging of the W\,48A \hiir\ and its surroundings would be necessary to trace the PDR kinematics. Finally, \citet{roshi:2005} found that the \hiir\ is receding with a few ($\sim$3) \,\kms , which might explain why the arc is receding rather than advancing with respect to the LSR velocity of the ridge. In Fig.~ \ref{fig:sketch}, we modify the view of \citet{roshi:2005} to reflect the results of our observations.

\subsection{Comparison of age estimates}
\label{disc:ages}
In Section \ref{sec:compact_sources} we derived the ages of the clumps/cores based on the $L/M$ tracks of \citet{molinari:2008} to be: $\sim$$8\times10^5$\,yr for clump H-1 (\hiir) and core H-2b (maser core),  and $\sim$$1.5\times10^5$\,yr for core H-3b (containing the molecular outflow). These ages contain many uncertainties: the uncertainties of the envelope masses and bolometric luminosities, and those from stellar evolution models used for the stellar tracks. It is therefore useful to compare these age estimations with the results from radio continuum data, molecular line data, and the literature.

Radio continuum and RRL measurements (\citealt{wood:1989a,onello:1994,roshi:2005}) unmistakably showed that the object matching with clump H-1 is a \hiir . Literature ages of \hiir s are similar to our age estimate: $6\times10^5$\,yr (\citealt{motte:2007}), $\sim10^5$\,yr (\citealt{wood:1989a,akeson:1996}). 

Apart from water and hydroxyl masers, core H-2a contains several class II methanol maser transitions, indicating the presence of a high-mass star in formation (\citealt{breen:2013}). Methanol masers are thought to be excited before the onset of the H{\sc ii} region and be quenched by the ionising radiation of the star (during the H{\sc ii} region phase), and should hence have younger or similar ages as H{\sc ii} regions, i.e., $\lesssim10^5$\, yr (\citealt{walsh:1998,codella:2000,breen:2010}). The detection of \mecn, a molecule typical of intermediate to high-mass protostellar objects such as hot cores (e.g, \citealt{olmi:1996,araya:2005,purcell:2006}), and the absence of a compact radio source (due to the free-free emission of the ionising star, which is typical of H{\sc ii} regions) also agrees with a younger evolutionary stage than a \hiir\ or a deeply embedded \hiir . Higher sensitivity radio observations could test if the object might possibly be a hyper compact H{\sc ii} region.  

Core H-3b was found to contain a high-mass young stellar object identified by the massive collimated outflow observed in CO\,(3--2) emission (Section~\ref{sec:outflow}) and HCO$^+$(1--0) line (\citealt{pillai:2011}). We detected \amm\ and \mecn\ emission and obtained the [\mecn]/[\amm] abundance ratio, which can be used as a chemical clock (\citealt{charnley:1992}). The abundance ratio found toward core H-3b was $\sim$1$\times10^{-3}$ (see Tables \ref{tab:ch3cn} and \ref{tab:nh3}), which is similar to the typical hot core abundance ratio of [\mecn]/[\amm] $\sim10^{-3}$ that corresponds to ages of $\sim$6$\times10^4$ yr (\citealt{charnley:1992}). Literature ages for hot cores (high-mass protostellar objects) are between $2\times10^4$\,yr (\citealt{motte:2007}) and $5\times10^4$\,yr (\citealt{cazaux:2003}). These two age estimates are both lower than what we obtained from the $L/M$ diagram.

Using several molecular lines and radio continuum data, we could infer that: 1) clump H-1 is indeed the most evolved and oldest object, 2) core H-2a is in an earlier evolutionary stage than clump H-1 and most likely also younger than clump H-1, while the $L/M$ diagram based age estimation found similar ages for these two objects, and 3) that core H-3b is indeed the youngest and less evolved object. Given that all three objects are forming high-mass stars the evolutionary difference is likely to reflect the age difference, since high-mass stars evolve along similar tracks, i.e. change their envelope mass and luminosity on similar timescales. It is thus important to have higher-resolution submm or radio continuum data and molecular line data to complement the age estimations based on infrared continuum data (such as the $L/M$ diagram).

\begin{figure*}
\includegraphics[angle=-90,width=13cm]{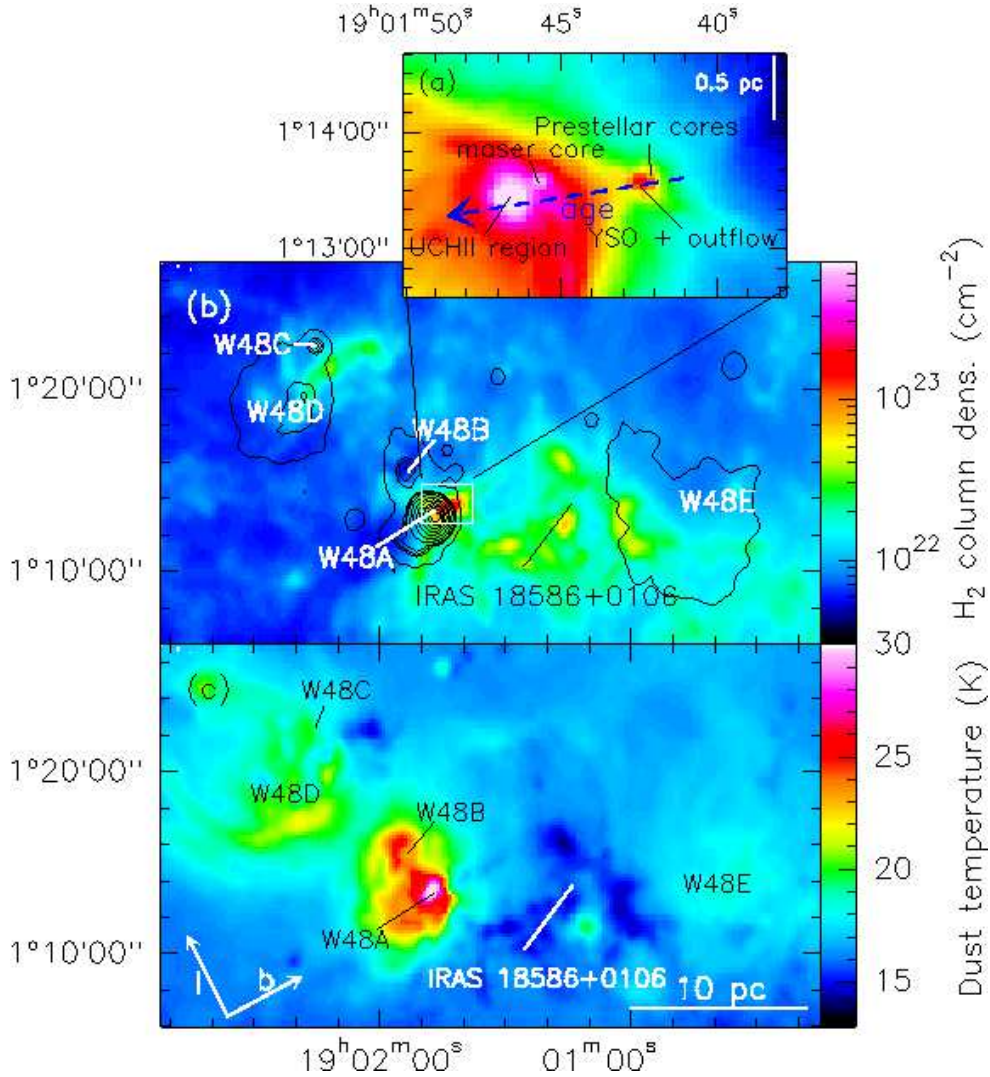}
\caption{\label{fig:overview} {\em Panel ($a$):} Sequential star formation around W\,48A indicated on the PACS 70\,$\mu$m map. {\em Panels ($b$) and ($c$):} {\em Herschel} H$_2$ column density and temperature maps of the W\,48 H{\sc ii} region complex (W\,48A--E). The column density map is overlaid with the 21\,cm contours of the NVSS survey (\citealt{condon:1998}). The Aquila supershell  expansion is oriented in the direction of Galactic latitude, $b$, indicated in the bottom panel.}
\end{figure*}

\subsection{Sequential star formation sequence around W\,48A?}

The top panel in Fig.~\ref{fig:overview} shows a synthesis of our sequential star formation results in W\,48A. In this study we isolated at least four different stages of star formation and estimated their ages, the \hiir\ (clump H-1), the maser core (core H-2a), the active star-forming YSO with outflow (core H-3b), and the cold streamers in which \citet{pillai:2011} found prestellar cores in NH$_2$D of 50--200\,$M_\odot$. The ages of prestellar objects depend strongly on whether they will form low- or high-mass stars. Assuming that these massive prestellar cores will form high-mass stars, then they should have younger ages than hot cores $<5\times10^4$yr (\citealt{cazaux:2003}). As protostellar heating destroys NH$_2$D, detecting these prestellar cores indicates that no protostars have been formed yet, which was confirmed through the non-detection of protostellar objects in the 70\,$\mu$m PACS map.
Interestingly, these four stages of star formation are aligned in a east (old) to west (young) configuration, forming a linear age gradient (Fig.~\ref{fig:overview}, panel ($a$)). Large projection effects can, most likely, be disregarded since the objects are forming within the same molecular cloud, suggesting that this remarkable age arrangement is real and reflects the formation history of the W\,48 molecular cloud.
 
Panels ($b$) and ($c$) in Fig.~\ref{fig:overview} show that the \hiir\ formed on the border of the molecular cloud. Core H-2a could have either formed before the creation of the arc and was swept up by it, or formed in the material collected by the PDR in a collect and collapse mechanism (\citealt{elmegreen:1977,whitworth:1994}). 
It is unlikely that the \hiir\ triggered the formation of the W\,48A molecular cloud nor the star formation inside it. In the previous section we mentioned the \hiir\ is confined by the PDR (\citealt{roshi:2005}), but that the PDR is slowly expanding into the molecular medium. The distance between \hiir\ and clump H-3 is about 1\,pc. A shock wave propagating at  $\sim$0.3\,\kms\ , which is the sound speed of the medium at 25\,K, would take about 3\,Myr. The age clump H-3 is about an order of magnitude younger than that, ruling out the \hiir\ (and the PDR) as the triggering agent. Hence, the W\,48A molecular cloud and its star-forming seeds were assembled before the onset of the \hiir\ in such a way that star formation started in the east and progressed toward the west. The sequential star formation in W\,48A reminds of the age gradient seen in the DR\,21 molecular cloud which was formed by colliding flows, and has a south (old) - north (young) age gradient (see \citealt{csengeri:2011b,hennemann:2012}). Given the location of W\,48 at almost 100\,pc below the Galactic plane, the presence of an age gradient suggests a large scale external force.

First of all, the location of the other W\,48 H{\sc ii} regions cannot explain the age gradient (by e.g., triggering) and it seems more likely that all these H{\sc ii} regions belong to one larger star-forming complex. Based on 21\,cm HI line data,  \citet{maciejewski:1996} discovered the Aquila supershell, centred at about four degrees below W\,48A, at $l$=35$^\circ$, $b$=--6$^\circ$. This shell is expanding from below into the Galactic plane, creating, in addition to a super bubble, a cone-like shape directed at $l$=$34\rlap{.}^\circ6$, $b$=$-1\rlap{.}^\circ4$, very near ($\sim$40\arcmin\ or $\sim$38\,pc) to the W\,48 H{\sc ii} regions. The 21\,cm HI spectra of the left and centre part of this cone show a clear peak at 40--45\,\kms (Figure 4, \citealt{maciejewski:1996}), which overlaps with the LSR velocity of the molecular material of W\,48A and of the other H{\sc ii} regions in W\,48: 41.2\,\kms\ for W\,48B; 46.7\,\kms\ for W\,48D; 45.5\,\kms\ for W\,48E (\citealt{onello:1994}). It is therefore likely that the initial molecular cloud in which the W\,48 complex of  H{\sc ii} regions was formed through the compression of ambient material by the Aquila supershell. \citet{maciejewski:1996} estimated that the total energy of the events creating the Aquila supershell are about $1-5\times10^{52}$\,erg, corresponding to 10--100 supernovae explosions powering the system over $10^7$ yr. The W\,48 H{\sc ii} regions were formed recently given the supershell's age ($\sim2\times10^7$\,yr) and its estimated expansion velocity of about 15\,\kms\ (which is variable depending on the density of the material it passes through). The distance between the centre of the supershell and the W\,48 H{\sc ii} regions is about 4 degrees (230\,pc at a distance of 3.27\,kpc) and at a velocity of 15\,\kms\ it would take about 1.5 Myr for the shell to reach the W\,48 H{\sc ii} regions, leaving ample time for the W\,48 H{\sc ii} regions to evolve into their current state. 

At the location of the W\,48 H{\sc ii} regions, the supershell's orientation is toward increasing Galactic latitudes. The shell would have therefore first reached the locations of W\,48C and D, then W\,48A and B, and finally W\,48E. Hence, the density structures created by the shell, should be older in the eastern side of Fig~\ref{fig:overview} and younger in the western side. Most of these H{\sc ii} regions are too evolved to have strong far-infrared emission, hence it is difficult to estimate their evolutionary stages from their envelope masses and bolometric luminosities. It is more fruitful to look at the size of the ionised  hydrogen region (the Str\"omgren shell) created by the young star, which one can observe through its free-free emission in the centimetre radio continuum. For this purpose we used the 21\,cm maps from the NRAO's NVSS survey (\citealt{condon:1998}). With AIPS task JMFIT we fitted 2D Gaussians to the 21\,cm emission of W\,48A--D obtaining the size, peak flux and integrated flux (Table \ref{tab:21cm}). The ratio of the integrated to peak flux (I/P), listed in the fifth column of Table~\ref{tab:21cm}, is a measure of the source compactness (1.0 for a point source and increasing when the source is more extended). W\,48D seems indeed to be an old H{\sc ii} region, since its radio emission very extended and it has possibly triggered another younger and compact H{\sc ii} region, W\,48C, which is very compact, hence very young, on one of its borders. The radio emission of W\,48B is less extended than W\,48D, but more extended that that of W\,48A, implying an intermediate evolutionary stage. This is consistent with W\,48B already containing a cluster of stars (clearly visible in the near-IR) and being surrounded by a small circular dust arc. Based on the 21\,cm radio continuum we conclude that the H{\sc ii} regions have an evolutionary gradient along Galactic latitude, such as one would expect if these regions were formed by the Aquila supershell.  In this discussion we have neglected W\,48E, whose nature, either a H{\sc ii} region or a supernova remnant, is uncertain, since it emits extended weak radio emission, and does not contain many YSOs, nor high dust column densities. 

\begin{table}
\begin{flushleft}
\caption{21\,cm emission properties of the W\,48 H{\sc ii} regions}
\label{tab:21cm}
\begin{tabular}{l c c c c}
\hline
H{\sc ii} region & Size & Peak Flux & Int. Flux & I/P\\
                           & (arcsec, arcsec) & (Jy~beam$^{-1}$) & (Jy) &\\
\hline
W\,48A & 74, 72 & 4.9 & 12.8 & 2.6\\
W\,48B & 118, 84 & 0.12 & 0.6 & 5.0\\
W\,48C & 50, 49 & 0.16 & 0.2 & 1.3\\
W\,48D & 221, 169 & 0.097 & 1.8 & 185.5\\
\hline
\end{tabular}

NOTES. Columns are (from left to right): name; size in major and minor axis at FWHM; Peak flux; Integrated Flux; ratio of Integrated to Peak flux.

\end{flushleft}
\end{table}

In addition to the W\,48 H{\sc ii} regions, IRAS 18586+0106, also known as Mol\,87, is located between W\,48A and W\,48E. IRAS 18586 contains a few massive star-forming clumps (\citealt{beltran:2006aa}),  but no centimetre continuum emission, which would indicate free-free emission of young stars, could be associated with them (\citealt{molinari:1998a}), nor methanol or water masers (\citealt{fontani:2010}). With its \amm\,(1,1) LSR velocity of 38\,\kms\ (\citealt{molinari:1996}), IRAS 18586 is very likely to be a part of the W\,48 complex. In Appendix~\ref{sec:iras} we analysed the infrared emission of the two clumps found in the {\em Herschel} maps. Source A, which coincides with the IRAS-coordinates centre, is the most evolved source with an age of $\sim$5$\times$10$^5$\,yr forming an intermediate to high-mass star. The second source (B) is colder, more massive and younger. For the beginning of star formation in  IRAS 18586, we only need to take the estimate of the oldest object into consideration, which is $\sim$5$\times$10$^5$\,yr. This is younger than most evolved object in W48\,A (the \hiir), and would agree with the predicted age gradient if all these star-forming regions were formed by the cone of the Aquila supershell. 

In addition, the {\em Herschel} column density and temperature maps (Fig.~\ref{fig:overview}) revealed that there exists a large reservoir of dense and cold gas west of W\,48A, implying that this region has a high potential to form stars. The morphology of this dense gas suggests that it has not been strongly affected by H{\sc ii} regions or stellar winds, that are known to shape the material into bubbles and shells. East of W\,48A, there is much less dense material, and when present it is shaped in shells, bubbles and ridges by H{\sc ii} regions. Hence, the large scale {\em Herschel} data support the hypothesis of an east-west evolutionary gradient across the W\,48 H{\sc ii} regions, in a roughly similar orientation as the age gradient found around W\,48A.  

\section{Summary}

We have presented a combined approach of high-sensitivity {\em Herschel} continuum observations with high-resolution molecular line data from the BIMA array to study the sequential star formation around the W\,48A \hiir . For the three clumps found with {\em Herschel}, which include a \hiir , a maser emitting YSO, and a YSO with a massive molecular outflow, we combined age estimations based on envelope mass and luminosity with those from radio continuum, molecular line data and the literature values, and found the ages to agree within an order of magnitude. Confusion of emission and multiplicity affect the $L/M$-based ages, by moving the star-forming object to more massive stellar tracks and to higher luminosities. We tried to correct for this by rescaling the continuum fluxes of clump H-2 and H-3 to the core-sized objects detected in the BIMA \mecn\ maps. High resolution continuum observations with ALMA, SMA or CCAT are necessary to reach the fragmentation limit of the cores for obtaining more accurate masses and luminosities. In addition to dust continuum data, radio continuum and molecular line data are crucial to identify the various evolutionary phases and to verify the ages based on envelope masses and luminosities. 

The W\,48A molecular cloud has a particular shape: it is concentrated along a ridge (the main axis of the cloud) lying in front of an arc (which, from our perspective is oriented toward the cloud) that partially surrounds the \hiir . \c18o\,(1--0) line data shows that this arc-like structure has two velocity components, the northern part which is receding and the southern part which is at rest/advancing. This arc, or part of it, was most likely created by the \hiir, when this formed at the eastern edge of the ridge, and is maintained by the over-pressured PDR. Based on the age dating of the {\em Herschel} sources, we found a linear evolutionary gradient across the W\,48A star-forming region that we interpreted as an age gradient. This age gradient starts in the east of W\,48A with the \hiir , the oldest object, and traverses the arc following the major axis of the W\,48A molecular cloud. Since the \hiir\ is pressure confined by the PDR, it could not have formed the W\,48A molecular cloud nor triggered the formation of most of the YSOs inside of it. 

From the centimetre radio continuum we found that a similar east (older) - west (younger) orientation, as in W\,48A is also present across the W\,48 H{\sc ii} regions. Also, large scale dust maps show that in the eastern part of W48 the H{\sc ii} regions shaped all the gas and dust around them (more evolved, older), while in the western part we find mostly molecular clouds (less evolved, younger). We therefore considered even larger scales and found that the Aquila supershell expansion is a plausible candidate to have swept up the material to form the W\,48 complex, since the velocity of the shell matches the LSR velocities of the W\,48A-E H{\sc ii} regions, and the age gradient of the W\,48 H{\sc ii} regions is roughly aligned with the direction of the expansion of the shell. 

\section*{Acknowledgments}
The authors thank the anonymous referee for his/her detailed comments which lead to a significant improvement of the manuscript.
Furthermore, the authors are very grateful to Gemma Busquet, Miguel Pereira-Santaella, and Scige J. Liu. The staff of both the BIMA and IRAM 30m telescopes are acknowledged for their support during the observations. KLJR acknowledges funding by the Agenzia Spaziale Italiana (ASI) under contract number I/005/11/0. DP is funded through the Operational Program ``Education and Life-long LearningÓ and is co-financed by the European Union (European Social Fund) and Greek national funds. Part of this work was supported by the French National Agency for Research (ANR) project ``PROBeS", number ANR-08-BLAN-0241. 

SPIRE has been developed by a consortium of institutes led by Cardiff University (UK) and including Univ. Lethbridge (Canada); NAOC (China); CEA, LAM (France); IFSI, Univ. Padua (Italy); IAC (Spain); Stockholm Observatory (Sweden); Imperial College London, RAL, UCL-MSSL, UKATC, Univ. Sussex (UK); and Caltech, JPL, NHSC, Univ. Colorado (USA). This development has been supported by national funding agencies: CSA (Canada); NAOC (China); CEA, CNES, CNRS (France); ASI (Italy); MCINN (Spain); SNSB (Sweden); STFC (UK); and NASA (USA).
PACS has been developed by a consortium of institutes led by MPE (Germany) and including UVIE (Austria); KU Leuven, CSL, IMEC (Belgium); CEA, LAM (France); MPIA (Germany); INAF-IFSI/OAA/OAP/OAT, LENS, SISSA (Italy); IAC (Spain). This development has been supported by the funding agencies BMVIT (Austria), ESA-PRODEX (Belgium), CEA/CNES (France), DLR (Germany), ASI/INAF (Italy), and CICYT/MCYT (Spain). The BIMA array was operated with support from the National Science Foundation under grants AST-9981308 to UC Berkeley, AST-9981363 to U. Illinois, and AST-9981289 to U. Maryland. The James Clerk Maxwell Telescope is operated by the Joint Astronomy Centre on behalf of the Science and Technology Facilities Council of the United Kingdom, the Netherlands Organisation for Scientific Research, and the National Research Council of Canada. This research has made use of the SIMBAD database and the VizieR service, operated at CDS, Strasbourg, France.
\bibliographystyle{aa} 
\bibliography{/Users/kazi/tot} 

\newpage
\appendix

\section{MAMBO 1.3\,mm map}
\begin{figure*}
\centering
\includegraphics[angle=-90,width=18cm]{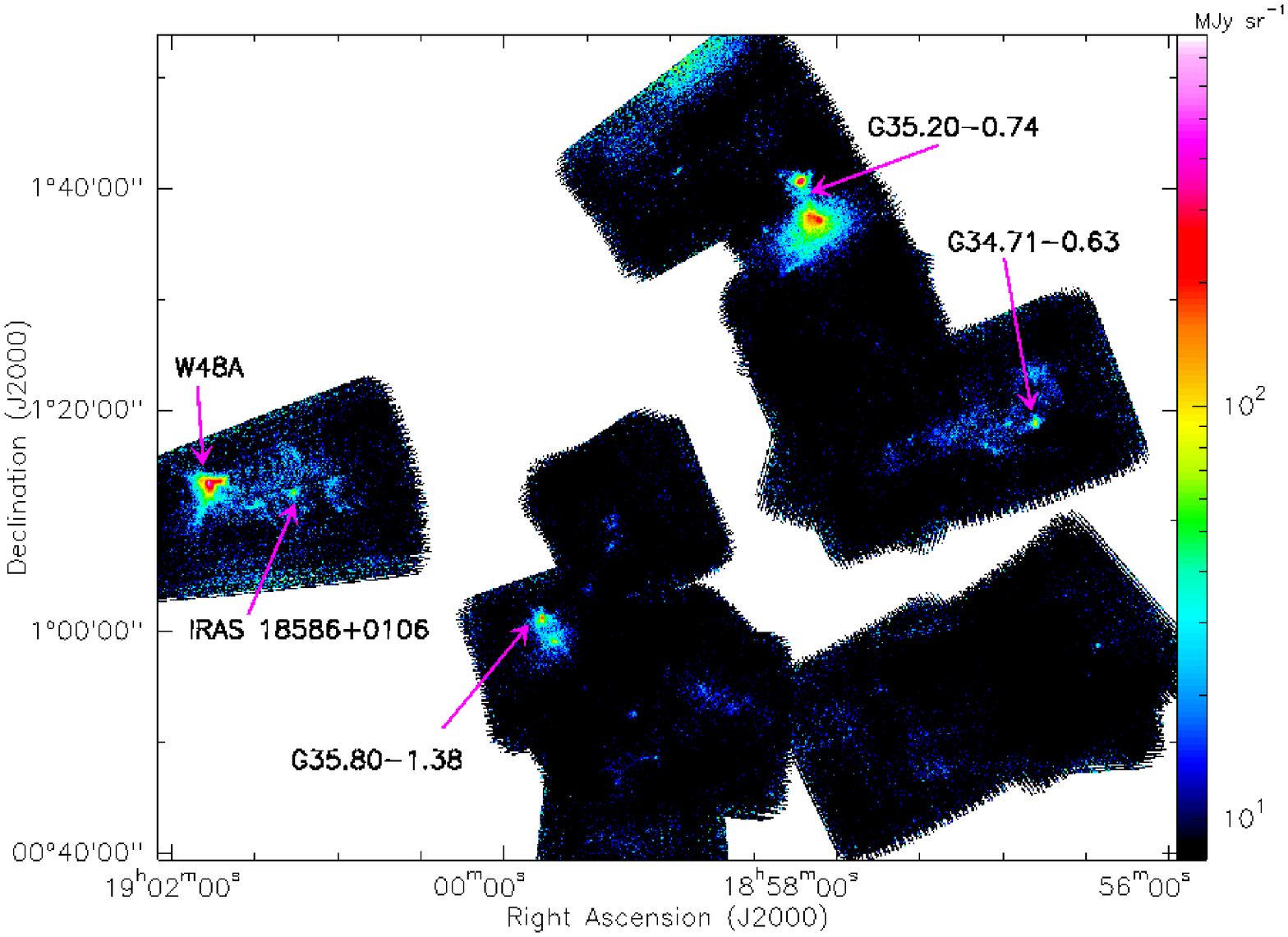}
\caption{\label{fig:mambo_allcov} Full MAMBO 1.3\,mm continuum map. Apart from W\,48A and IRAS 18586, other mm-bright star-forming regions in the field are indicated.}
\end{figure*}

The 1.3\,mm MAMBO data of W\,48A shown in Fig.~\ref{fig:continuum_all} were part of a larger map presented in Fig.~\ref{fig:mambo_allcov}. Observational details are given in Section~\ref{sec:30m} and Table~\ref{ta:spec}.

\section{Distribution of 107\,GHz methanol masers}
\label{sec:maser_app}

\begin{figure}
\includegraphics[angle=-90,width=8cm]{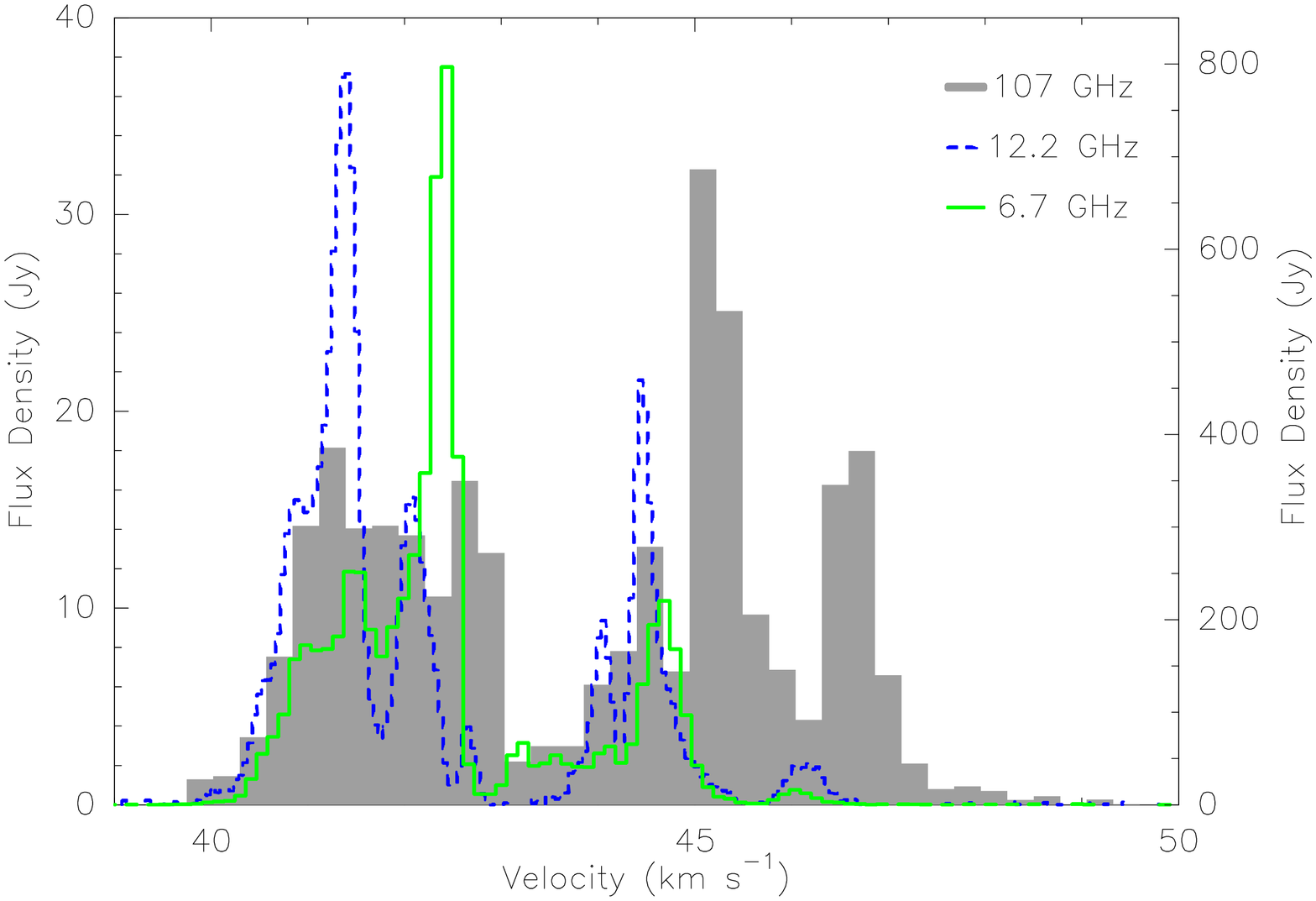}
\caption{\label{fig:maser_spec} Spectra of 107\,GHz (grey histogram), 12.2\,GHz (dashed blue), and 6.7\,GHz (green) methanol masers toward the clump H-2 in W\,48A. The left y-axis is for the 107 and 12.2\,GHz transitions, while the right y-axis belongs to the 6.7 GHz methanol maser.}
\end{figure}

The spectra of the 107-GHz masers compared to the 12.2-GHz (M. Gaylard, priv. comm.) and 6.7-GHz masers (\citealt{goedhart:2004}) are shown in Fig.~\ref{fig:maser_spec}. While the 6.7- and 12.2- GHz masers share the same velocity range of 39.5--46\,$\mathrm{km~s^{-1}}$, the 107 GHz masers show emission out to 48\,$\mathrm{km~s^{-1}}$, indicating that various transitions probe similar, but not the same physical conditions. The relative intensities of individual maser features are different at each transition, with the red-shifted features brighter at 107 GHz (see also the work of \citealt{valtts:2002}).

\begin{figure}
\resizebox{\hsize}{!}{\includegraphics[clip,angle=-90]{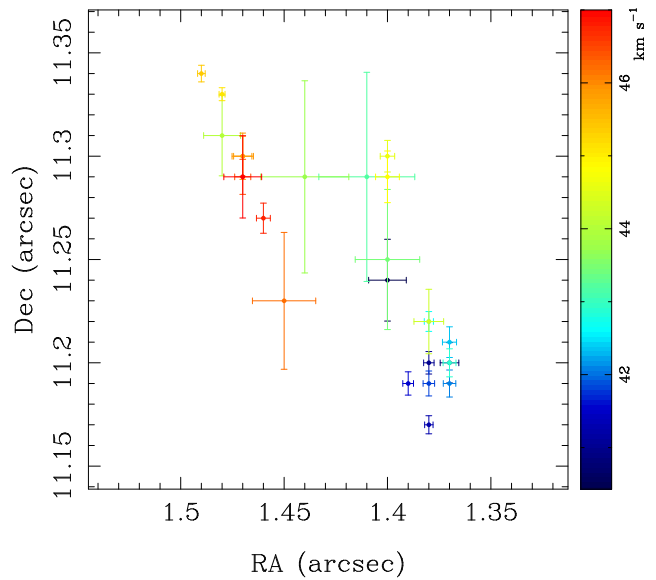}}
\caption{\label{fig:spotmap} 107\,GHz methanol maser map obtained through gaussian fitting. Weaker features have higher positional uncertainties. Axes give the offset in arcseconds with respect to the pointing centre 19$^\mathrm{h}$01$^\mathrm{m}$45$\rlap{.}^\mathrm{s}$448 +01\degr13\arcmin21$\rlap{.}$\arcsec493 (J2000). }
\end{figure}

The 107\,GHz methanal maser spots were detected across a velocity range of $\sim$8\,\kms\ (see Fig.~\ref{fig:maser_spec}) and appear to be unresolved in the data cubes generated. We obtained the positions of the methanol maser spots by fitting two-dimensional Gaussians to the data cube from the (highest resolution) BIMA A-array data. The beam-size from A-array data alone is 0$\rlap{.}$\arcsec 57$\times$0$\rlap{.}$\arcsec24 (Table \ref{ta:spec}) but two-dimensional Gaussian fitting can find the centre of the maser spot to much higher accuracy, $\sim$0$\rlap{.}$\arcsec 01(33\,AU). The spatial and velocity distribution of the 107\,GHz methanol masers is shown in Fig.~\ref{fig:spotmap}. 

Methanol maser kinematics can trace the small scale gas motions near the protostar, such as envelope expansion or a rotating disk (\citealt{moscadelli:2011}). AFGL\,5142, is an example of a high-mass protostar where the 3D kinematics of methanol masers seem to be emanating from a rotation disk (\citealt{goddi:2011}). The general morphology and angular distribution of the 107-GHz masers is similar to that found for the 6.7- and 12.2 GHz masers by \citet{minier:2000}: spots belonging to the red-shifted group in the north-east and spots belonging to the blue-shifted group in the south-west. However, a one-to-one correspondence of spots in the different transitions is not apparent. The 107 GHz emission appears to be coming from a smaller area than seen in \citet{minier:2000}: $\sim$0$\rlap{.}$\arcsec 21 compared to the $\sim$0$\rlap{.}$\arcsec 43. The 107\,GHz masers show a tighter correlation in position and velocities than the 6.7 and 12.2 GHz masers. We find a correlation coefficient of 0.86 by performing a linear fit to the 107 GHz maser distribution. If we assume Keplerian rotation and a distance of 3.27 kpc (\citealt{zhangb:2009}), the mass enclosed in 0$\rlap{.}$\arcsec21 (0.003\,pc or 685\,AU) is derived to be 2.2$\cos^{-2}(i)$\,$M_\odot$. Furthermore, if we interpret the spatial distribution of 6.7- and 12.2\,GHz masers in \citet{minier:2000} as due to the inclination $i$ of the disk, we obtain an inclination of 30\degr . Applying this inclination, the enclosed mass increases to 3\,$M_\odot$, which about a factor 2 lower than the mass obtained by \citet{minier:2000} enclosed in 1300\,AU. Class II methanol masers, to which the 6.7, 12.2 and 107\,GHz masers belong, are associated with {\em high-mass} star formation (\citealt{menten:1991,minier:2003,breen:2013}). Our derived mass of 3\,$M_\odot$ is unlikely to form a high-mass star, indicating that the 107\,GHz masers do not allow a mass calculation, because, for example, they do not trace the entire disk, as also suggested recently in \citet{breen:2013}.

\section{Distribution of \amm(2,2)/\amm(1,1) main beam temperatures}

\begin{figure}
\includegraphics[width=8cm]{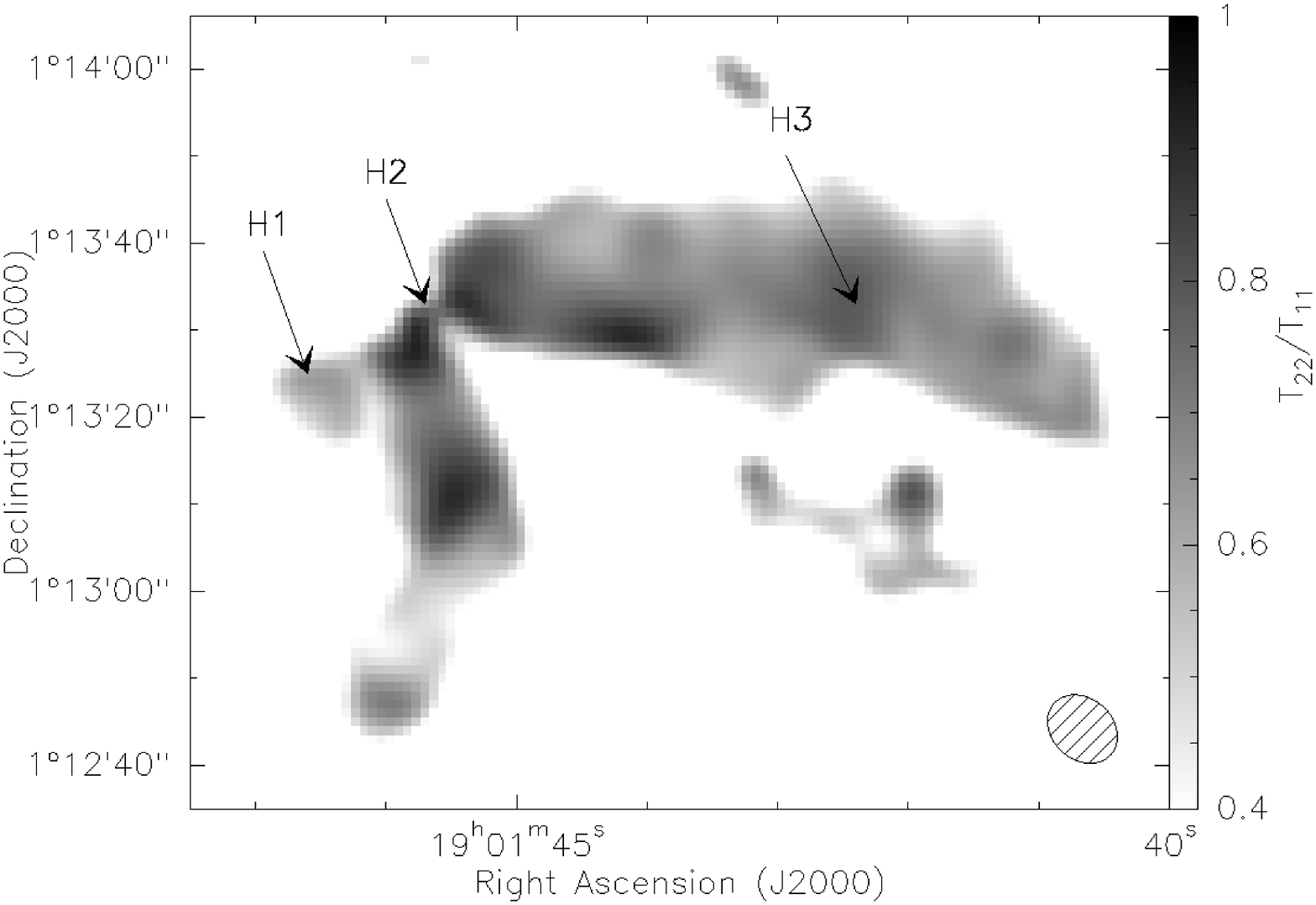}
\caption{\label{fig:ammratio} Ratio of \amm(2,2) and \amm(1,1) main beam temperatures. The beam is shown in the bottom right corner.}
\end{figure}

The ratio of the \amm(2,2) to \amm(1,1) main beam temperature is a sign of increasing kinetic temperature (\citealt{torrelles:1989,zhang:2002}), and can be used to trace heating sources. Figure \ref{fig:ammratio} shows the distribution of this ratio based on the VLA cubes. Near clump H-2 the ratio is high. This sign of heating is consistent with the presence of a PDR as seen in previous studies of PDRs near H{\sc ii} regions (e.g., \citealt{palau:2007}). Note that the gas near clump H1 has a low ratio, which indicates that this gas is cold, and hence in front of the \hiir\ -- this agrees with the \amm\ absorption seen at this location (see section \ref{sec:nh3}).

\section{IRAS 18586+0106}
\label{sec:iras}

\begin{figure*}
\includegraphics[angle=-90,width=18cm]{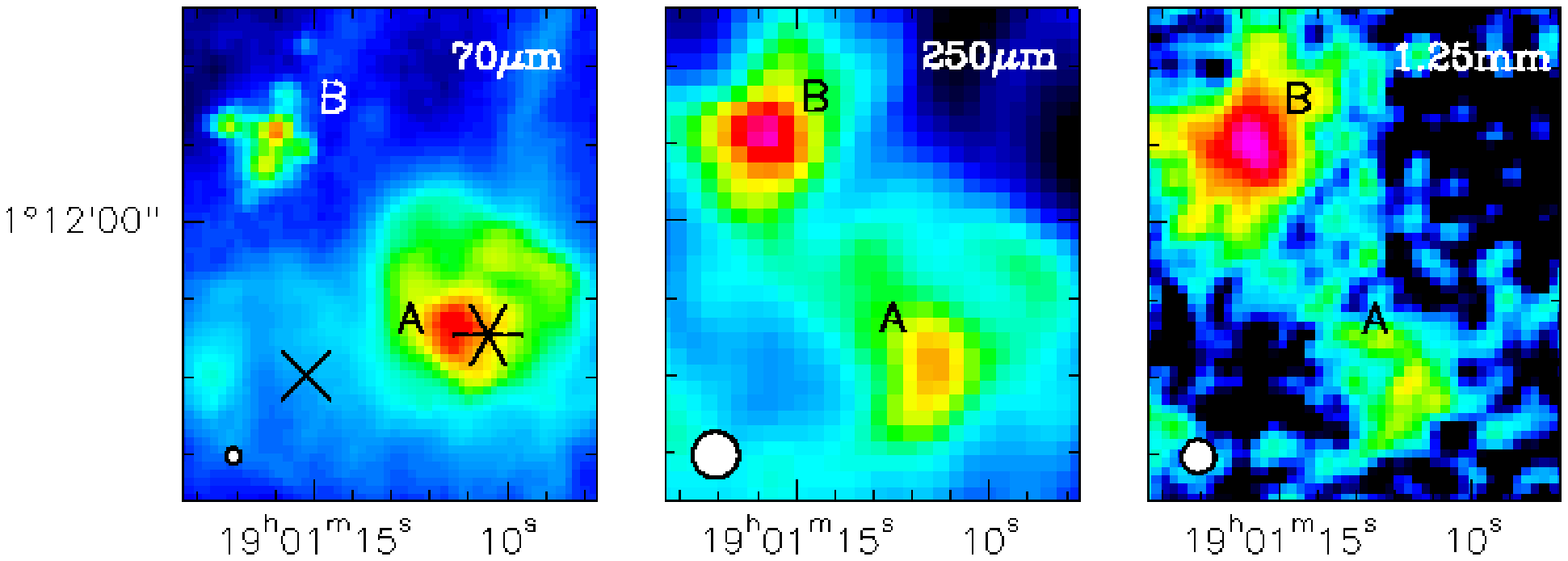}
\caption{\label{fig:iras_maps} {\em Herschel} 70\,$\mu$m, 250\,$\mu$m and MAMBO 1.2\,mm maps of IRAS 18586+0106 revealing two clear clumps, A and B. The position of the IRAS source is marked by a star. The radio source (\citealt{molinari:1998a}) is marked by a cross. The beam FWHM is given in the bottom left corner. }
\end{figure*}

\begin{figure}
\includegraphics[angle=-90,width=8cm]{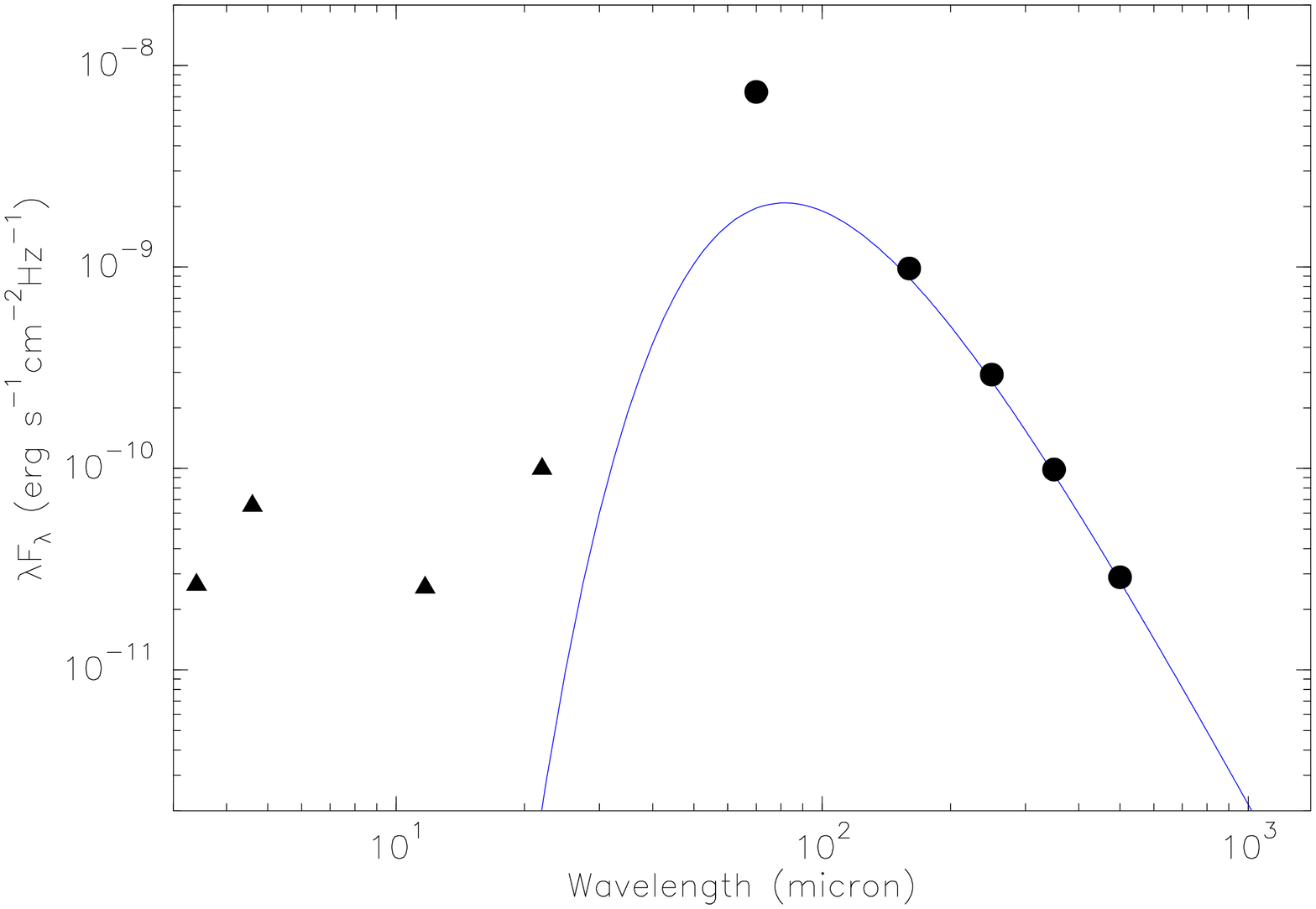}
\caption{\label{fig:iras_sed} Best modified black body fit to the IRAS 18586+0106 spectral energy distribution. WISE data points (\citealt{wrightwise:2010}) are marked with triangles, {\em Herschel} datapoints with filled dots. }
\end{figure}

IRAS 18586+0106 originated from the {\em IRAS}-colour-selected list of {\em low} sources by \citet{palla:1991}, which was consequently studied in \amm , centimetre and submillimetre emission (\citealt{molinari:1996,molinari:1998a,molinari:2000,beltran:2006aa}). The object was found to have weak \amm\ and submillimetre emission, and no radio emission (the radio emission was found to be uncorrelated with the {\em IRAS} source. The {\em IRAS} resolutions at the various wavelengths (for example, the point spread function (PSF) at 60\,$\mu$m was $\sim$2\arcmin$\times$5\arcmin) are nowadays surpassed by those of {\em Herschel} ($\sim$6\arcsec at 70\,$\mu$m). Figure \ref{fig:iras_maps} shows the 70 and 250\,$\mu$m {\em Herschel} and 1.2\,mm MAMBO maps of the {\em IRAS} source. Note that the image is completely within the {\em IRAS} PSF at 60\,$\mu$m, and that two clear clumps, marked as A and B are present. Of these, source A (which coincides with the centre of the {\em IRAS} source) is brighter at shorter wavelengths indicating a later evolutionary stage than that of source B, which is more luminous in the far-infrared to submillimetre. Performing the source extraction and SED fitting (Fig.~\ref{fig:iras_sed}), as described in Section~\ref{sec:compact_sources}, we find that source A has a temperature of 35\,K, a envelope mass of 70\,$M_\odot$ and a bolometric luminosity of 2140 $L_\odot$ (including the WISE data from \citealt{wrightwise:2010}). Source B (SED not shown) has a temperatures of 17\,K, mass of  300\,$M_\odot$, bolometric luminosity of 470\,$L_\mathrm{bol}$.  
  
Source A is located between the evolutionary tracks of the 6.2 and 8.0\,$M_\odot$ stars in the $L/M$ diagram (\citealt{molinari:2008}), suggesting it will form an intermediate to high mass star and that it has an age of $\sim$4.5$\times$10$^{5}$\,yr. Source B lies on the 13.5\,$M_\odot$ track and its combination of estimated mass and luminosity gives an age of  3$\times$10$^{5}$\,yr.

\bsp
\label{lastpage}
\end{document}